\documentclass{emulateapj}
\usepackage[breaklinks,colorlinks,urlcolor=blue,citecolor=blue,linkcolor=blue]{hyperref}
\usepackage{mathrsfs}
\usepackage{amsmath}
\usepackage{graphicx}
\usepackage{subfigure}
\usepackage{url}
\usepackage{hyperref}

\shortauthors{B.D. Metzger \& N.C. Stone}

\shorttitle{Periodic Flares from Colliding EMRIs}

\newcommand\degd{\ifmmode^{\circ}\!\!\!.\,\else$^{\circ}\!\!\!.\,$\fi}

\newcommand{\be}{\begin{equation}}
\newcommand{\ee}{\end{equation}}

\begin{document}

\title{Periodic accretion-powered flares from colliding EMRIs as TDE Imposters}

\author{
Brian D.~Metzger$^{1\dagger}$
\& Nicholas C.~Stone$^{1*}$
}

\altaffiltext{1}{Columbia Astrophysics Laboratory, Pupin Hall, New
  York, NY, 10027, USA.}
  \altaffiltext{$\dagger$}{bmetzger@phys.columbia.edu.}
\altaffiltext{*}{Einstein Fellow.}

\begin{abstract}
When a main sequence star undergoes Roche lobe overflow onto a supermassive black hole (SMBH) in a circular extreme mass ratio inspiral (EMRI), a phase of steady mass transfer ensues.  Over millions of years, the binary evolves to a period minimum before reversing course and migrating outwards as a brown dwarf.  Because the time interval between consecutive EMRIs is comparable to the mass-transfer timescale, the semi-major axes of two consecutive mass-transferring EMRIs will cross on a radial scale of less than a few AU.  We show that such EMRI crossing events are inevitably accompanied by a series of mildly relativistic, grazing physical collisions between the stars.  Each collision strips a small quantity of mass, primarily from the more massive star, which generally increases their radial separation to set up the next collision after a delay of decades to centuries (or longer) set by further gravitational radiation.  Depending on the mass of the SMBH, this interaction can result in $N_{\rm c} \sim 1-10^{4}$ gas production events of mass $\sim M_{\odot}/N_{\rm c}$, thus powering a quasi-periodic sequence of SMBH accretion-powered flares over a total duration of thousands of years or longer.  Although the EMRI rate is 2-3 orders of magnitude lower than the rate of tidal disruption events (TDE), the ability of a single interacting EMRI pair to produce a large number of luminous flares - and to make more judicious use of the available stellar fuel - could make their observed rate competitive with the TDE rate, enabling them to masquerade as ``TDE Imposters."  Gas produced by EMRI collisions is easier to circularize than the highly eccentric debris streams produced in TDEs.  We predict flares with bolometric luminosities that decay both as power laws shallower than $t^{-5/3}$ or as decaying exponentials in time.  Viscous spreading of the gaseous disks produced by the accumulation of previous mass-stripping events will place a substantial mass of gas on radial scales $\gtrsim 10-100$ AU at the time of a given flare, providing a possible explanation for the ``reprocessing blanket" required to explain the unexpectedly high optical luminosities of some candidate TDE flares.        
 \end{abstract}


\section{Introduction}

Compact star clusters around supermassive black holes (SMBHs) in galactic nuclei are the densest known stellar systems in the Universe.  Rapid exchange of angular momentum between the stars and stellar remnants in these environments causes stars to be occasionally perturbed onto low angular momentum (``loss cone'') orbits, which bring them near or within the tidal radius of the SMBH.

The vast majority of ill-fated stars descend to the SMBH on nearly parabolic orbits, from initial distances comparable to the parsec-scale SMBH sphere of influence (e.g. \citealt{Magorrian&Tremaine99,Wang&Merritt04,Stone&Metzger16}).  Depending on the depth of the pericenter distance relative to the tidal radius, these ``plunge" events result in a partial or complete dynamical disruption of the star accompanied by strong tidal compression, a phenomenon commonly known as a ``tidal disruption event" (TDE; \citealt{Hills75,Carter&Luminet83,Evans&Kochanek89,Lodato+09,Guillochon&RamirezRuiz13,Stone+13}).  
Prompt accretion by the SMBH of the gaseous debris of the disrupted star following a TDE was predicted to power a luminous flare (\citealt{Rees88,Phinney89,Ulmer99}), characterized by a rise time of weeks to months and followed by a $\propto t^{-5/3}$ power-law decay in the bolometric light curve set by the declining rate of mass fallback at late times.  

Despite this initial picture, recent numerical and analytic works have demonstrated that it is non-trivial to circularize the highly eccentric stellar debris streams created by the TDE into the compact accretion disk needed to power a flare (e.g., \citealt{Hayasaki+13,Shiokawa+15,Dai+15,Hayasaki+16,Guillochon&RamirezRuiz15,Bonnerot+16,Coughlin+16,Sadowski+16,Bonnerot+17,Tejada+17}).  This has lead to speculation that only a small fraction of TDE are accompanied by luminous flares, with most instead being ``dark" (\citealt{Guillochon&RamirezRuiz15,Hayasaki+16}) or producing emission through less efficient mechanisms than standard thin-disk accretion (\citealt{Shiokawa+15,Piran+15,Miller15,Metzger&Stone16}).  This possibility is supported by the discrepency between the theoretically-predicted TDE rate (\citealt{Stone&Metzger16,Kochanek16}) and the lower flare rate observed from optical surveys (e.g., \citealt{vanVelzen+11,vanVelzen&Farrar14, Holoien+16}).  
% This tension was further exacerbated by the discovery that TDEs preferentially occur in a rare subset of galaxies (\citealt{Arcavi+14,French+16}), thereby increasing the rate discrepency in normal galaxies.  Some have suggested that nuclear transients claimed to be TDE flares are not TDEs at all (\citealt{Saxton+16}), motivating a reevaluation of the types of transients expected to inhabit dense stellar systems surrounding SMBHs.   

In addition to those stars on parabolic plunges fated to produce TDEs, a smaller fraction approach the SMBH on more tightly bound orbits with lower eccentricities.  These ``extreme mass ratio inspirals" (EMRIs) have received attention as gravitational wave sources for the Laser Interferometer Space Antenna (LISA; \citealt{LISA17}).  This is especially true when the EMRI bodies are compact remnants (white dwarfs, neutron stars, and stellar mass black holes) for which tidal forces play little to no role in their inspiral evolution prior to their final plunge into the event horizon (although see \citealt{Zalamea+10}).

EMRIs of {\it main sequence} stars have received comparatively less attention than their compact remnants, in part because matter interactions alter the gravitational wave inspiral signal  (\citealt{Linial&Sari17}) and thus reduce their potential as pristine probes of general relativity.  If a main sequence star approaches the tidal sphere on a nearly circular orbit and begins overflowing its Roche lobe, mass transfer onto the SMBH is stable, resulting in the star being slowly accreted over millions of years  \citep{King&Done93, Dai&Blandford13,Linial&Sari17}.  When the response of the star to mass loss is adiabatic, or once its equation of state becomes dominated by electron degeneracy after it loses sufficient mass, the radius of the star expands and its orbit evolves to {\it larger} semi-major axes as a the result of further mass loss.  The system may be described during this phase as an extreme mass ratio ``outspiral" (e.g., \citealt{Dai&Blandford13,Linial&Sari17}), similar to the orbital evolution of cataclysmic variables following the period minimum.

The rate of EMRIs due to bodies entering the loss cone via two-body gravitational interactions is estimated to be $\sim 1\times 10^{-6}$ yr$^{-1}$ per galaxy (\citealt{BarOr&Alexander16, Aharon&Perets16}; see also \citealt{Hils&Bender95,Sigurdsson&Rees97,Freitag01,Ivanov02,Alexander&Hopman03,Hopman&Alexander06b}), roughly $2$ orders of magnitude lower than the TDE rate.  Due to the effects of mass segregation in nuclear clusters, this rate is also dominated by stellar mass black holes instead of lower mass main sequence stars or white dwarfs (e.g.~\citealt{Hopman&Alexander06b,Aharon&Perets16}).  Although the eccentricities of these ``two-body" EMRIs are much lower than those of the plunge events, they are usually still significant ($e \gtrsim 0.5-0.9$ on scales of the tidal radius; \citealt{Hopman&Alexander05}), such that the fate of most main sequence EMRIs delivered by two-body interactions will also be tidal disruption, though their light curves will differ from canonical $1-e\ll 1$ TDEs \citep{Hayasaki+13}.

More nearly circular EMRIs are created by the tidal separation of stellar binaries by the SMBH (\citealt{Miller+05,AmaroSeoane+12}).  This is the same process hypothesized to produce the cluster of S-stars orbiting Sgr A$^{*}$ (e.g., \citealt{Perets+09}) and hypervelocity stars from our galactic centre  (e.g.~\citealt{Hills88,Sari+10}).  \citet{AmaroSeoane+12} estimate the rate of circular EMRIs to be $\sim 10^{-7}$ yr$^{-1}$ per galaxy for an assumed binary fraction of $5\%$ (lower than in the field due to the dissociation of the soft binaries in the dense stellar environment of the nuclear cluster).  This rate could be up to two orders of magnitude higher in galactic nuclei with large numbers of massive perturbers \citep{Perets+07} or nuclear spiral arms \citep{Hamers&Perets17}.

%This is compar the rate of higher eccentricity two-body main sequence EMRIs, once accounting for the reduction due to mass segregation in the latter.

This work develops an observable consequence of the existence of stable mass-transferring EMRIs, which is closely related to the fact that the lifetime of these systems is comparable to the average time interval between consecutive EMRIs.  As we will show, this naturally predicts a collision, or a series of quasi-periodic collisions, between every two inspiraling (or, more typically, one inspiraling and one outspiraling) stars as their semi-major axes cross on a radial scale of $\lesssim$ few AU.  Such collisions, which happen deep within the potential well of the black hole at relative velocities up to several tenths of the speed of light, can seriously damage or completely obliterate both stars, leading to punctuated episodes of sizable gas production.  

\begin{figure*}[!t]
\includegraphics[width=1.0\textwidth]{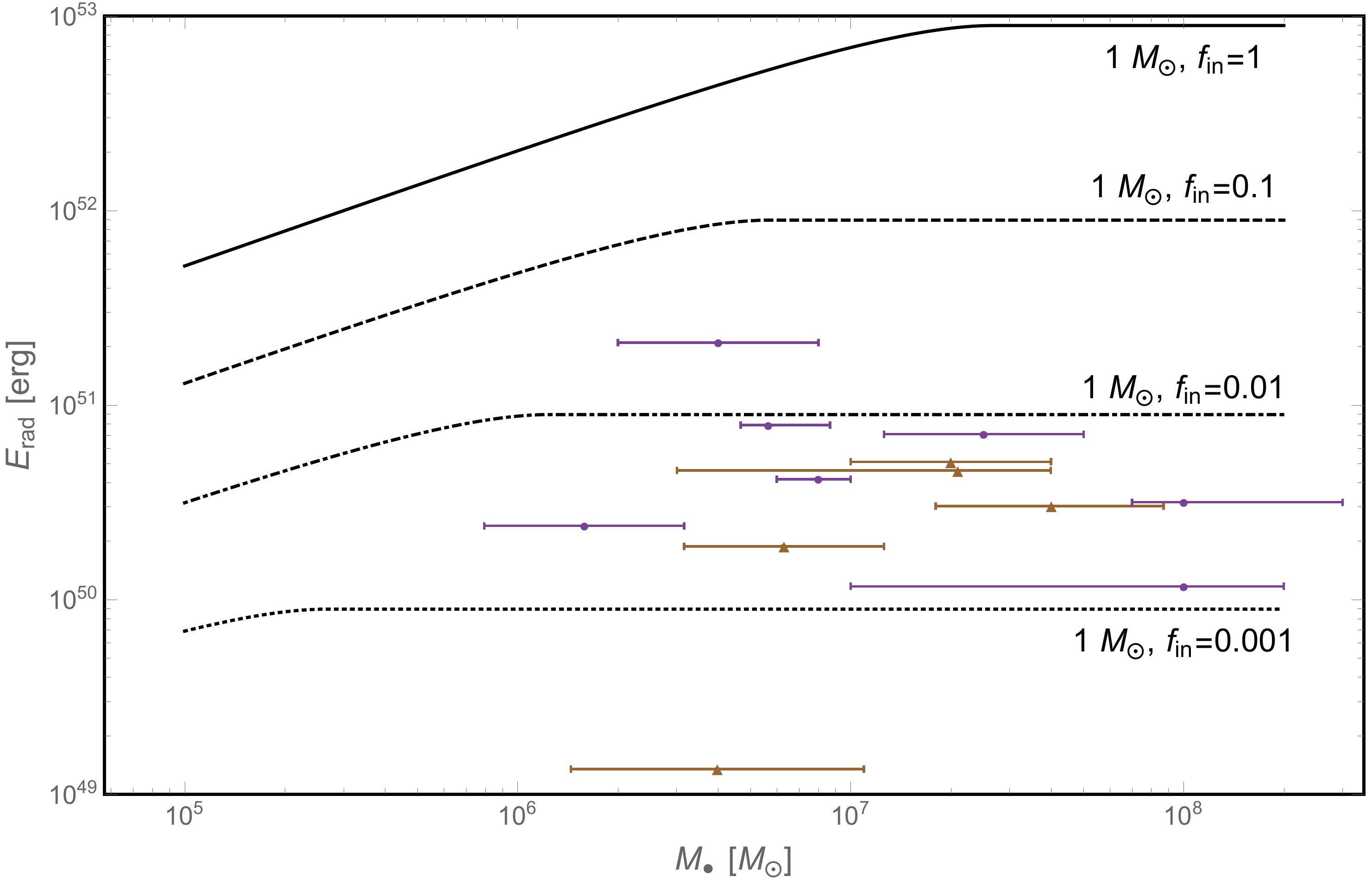}
\hspace{0.0cm}
%\includegraphics[width=0.35\textwidth]{schematic.pdf}
%\vspace{-0.4cm}
\caption{\footnotesize
{Total radiated energy $E_{\rm rad}$ of optically-selected TDE flare candidates as a function of SMBH mass, calculated using bolometric light curve data compiled by \citet{Hung+17} (estimated by fitting the optical/NUV SEDs to a single temperature blackbody), and using redshifts from the Open TDE Catalog\footnote{\url{http://TDE.space}}.  Purple circles show flares where the light curve is temporally resolved, while brown triangles show lower limits on $E_{\rm rad}$ in cases when the light curve peak is not resolved.  Overplot is the expected value of $E_{\rm rad} = 0.1 (f_{\rm in}M_{\star})c^{2}$ following the accretion of a $M_{\star} = 1M_{\odot}$ star for different accretion efficiencies $f_{\rm in}$, but limiting the radiated luminosity at all times to the Eddington luminosity for the given SMBH mass assuming a peak flare duration equal to the expected mass fall-back time.  This is an updated version of a similar plot (Fig. 2) from \citet{Stone&Metzger16}.  The 12 data points represent the TDE candidates iPTF16axa, ASASSN-14ae, PS1-11af, PS1-10jh, ASASSN-14li, iPTF16fnl, ASASSN-15oi, TDE1, TDE2, D1-9, D3-13, and PTF09ge.  Estimates for $M_\bullet$ and its error range are taken from \citet{Hung+17, Holoien+14, Chornock+14, Gezari+12, vanVelzen+16, Blagorodnova+17, Holoien+16b, vanVelzen+11, Gezari+08, Arcavi+14}.  In two cases (ASASSN-14ae and ASASSN-15oi) the discovery papers do not quote errors on SMBH mass estimates, so for these we have plotted approximate error intervals corresponding to 0.3 dex of scatter typical to galaxy scaling relations.}
}
\label{fig:Erad}
\end{figure*}

Rapid accretion of this gas by the SMBH powers a single luminous flare, or a series of periodic flares lasting hundreds to thousands of years or longer, caused by multiple grazing encounters between the stars separated by intervals of decades to centuries.  Although this emission mechanism is qualitatively similar to the common picture of TDE flares, EMRI collision flares may be observed more frequently than would be naively guessed from the low EMRI rate.  This is because standard TDEs may suffer from low radiative efficiencies due either to super-Eddington accretion rates \citep{Ayal+00, Metzger&Stone16} or, conversely,  difficulty circularizing highly eccentric debris streams \citep{Hayasaki+13, Shiokawa+15, Guillochon&RamirezRuiz15, Hayasaki+16, Bonnerot+16}.  Indeed, small accreted masses in putative TDE flares are consistent with the low bolometric energy fluences inferred by their optical/UV SEDs relative to the expection of thin disk accretion, as shown in Fig.~\ref{fig:Erad} (though the minority of flares seen to exhibit luminous IR dust echoes may have higher bolometric luminosities; \citealt{vanVelzen+16}).   Collisions among circularized EMRIs can make more judicious use of the available stellar mass budget, as accretion rates are generally sub-Eddington and collisionally liberated gas is produced on initially circular orbits.

%This is because circularization of the debris can be more rapid than in the eccentric disruption case, and because multiple outbursts can be produced by the same pair of crossing stars, resulting in a more judicious use of the star's accretion fuel, especially if the radiative efficiency of normal TDE flares is low.  

This scenario raises the possibility that some nuclear transients currently identified as TDE flares may in fact be EMRI collision products, i.e. ``TDE Imposters".  Indeed, we show that otherwise puzzling behavior seen in some of these flares, such as light curves which decay exponentially in time rather than as power-laws, may find a natural explanation in the EMRI collision model.  Viscous spreading of the accretion disks (out to radial scales of $\gtrsim 10-100$ AU) made in regular gas production events over the prolonged interval of the collisional flaring activity also provides an alternative source for the extended reprocessing layer needed to explain the unexpectedly high optical luminosities of many TDE flares.

This paper is organized as follows.  In $\S\ref{sec:emri}$ we review the evolution of stable mass transfer from a main sequence star onto a SMBH.  In $\S\ref{sec:collision}$ we discuss the conditions required for a collision between two consecutive EMRI stars.  In $\S\ref{sec:transient}$ we discuss the observable signatures of such a collision, including periodic flares from SMBH accretion, compare them to the observed population of TDE flares, and discuss the interactions of main sequence EMRIs with bona fide TDEs.  In $\S\ref{sec:conclusions}$ we briefly summarize our conclusions.

%Although the number density of stellar mass black holes and stars is predicted to be comparable on radial scales of $r \sim 0.01$ pc, the stellar density is predicted to rise less steeply ($n_{\star} \propto r^{-1.4}$) than the black hole density ($n_{\bullet} \propto  r^{-1.75}$) due to mass segregation (e.g.~\citealt{Antonini13}), such that on scales of $\sim 10^{-3}$ pc from which most of the EMRI loss cone flux originates (\citealt{BarOr&Alexander16}), the stellar density could be a few times lower than this prediction, i.e. $\sim 0.3-1\times 10^{-6}$ yr$^{-1}$ per galaxy.

\section{EMRI Mass Transfer Evolution}
\label{sec:emri}

%A star with a semi-major axis $a_0$ and eccentricity $e_0 \lesssim 1$ will see each decay in tandem due to gravitational radiation according to $a(t)/a_0 \simeq 2(1-e_0)(e/e_0)^{12/19}/(1-e^{2})$ (\citealt{Peters64}), such that an initially mild eccentricity $e_0 \sim 0.5$ will be reduced to $e \lesssim 10^{-4}$ as the semi-major axis reduces from milliparsec to AU scales from the SMBH.  For simplicity we hereafter assume circular orbits on the AU scales relevant to the mass transfer evolution.  

We consider stars inspiraling into the SMBH on nearly circular orbits when they reach the point of Roche lobe overflow (RLOF) on radial scales of a few AU (see below).  As discussed above, this will not be satisfied for most of the main sequence EMRIs produced by two-body scattering or resonant relaxation, which instead will possess high eccentricies $e \gtrsim 0.5$ at this separation and will likely undergo tidal disruption.  

However, EMRIs created by tidally detached binaries will generally possess lower eccentricities $e \sim 0.01-0.05$ when their pericenters reach these distances (\citealt{Miller+05,AmaroSeoane+12}).  Depending on the competition between circularization of the orbit due to gravitational wave emission versus that due to tidal friction, these stars will either undergo tidal disruption or they will end up in circular, stably mass-transferring orbits (see discussion in \citealt{AmaroSeoane+12}).  We focus on the latter case, which accounts for a large portion of the allowed parameter space for a reasonable range in the theoretically uncertain value of the $Q$ parameter controlling the rate of tidal circularization.

Once in a nearly circular orbit of semi-major axis $a$, a star of mass $M_{\star}$ orbiting the SMBH of mass $M_{\bullet} \gg M_{\star}$ loses energy to gravitational wave (GW) emission on the timescale
\be
\tau_{\rm GW} \equiv \frac{a}{|\dot{a}|} \simeq \frac{5}{64}\frac{ c^{5} a^{4}}{G^{3}M_{\star}M_\bullet^{2}} \approx 1.3\times 10^{4}\,{\rm yr}\,\,\frac{a_{\rm AU}^{4}}{M_{\star,\odot}M_{\bullet,7}^{2}},
\label{eq:tauGW}
\ee
to lowest post-Newtonian order.  Here $M_{\star,\odot} \equiv M_{\star}/M_{\odot}$, $M_{\bullet,7} \equiv M_{\bullet}/10^{7}M_{\odot}$, $a_{\rm AU} \equiv a/$AU, and we have written the gravitational constant and the speed of light as $G$ and $c$, respectively.

The size of the star's Roche lobe in the extreme mass ratio limit is given by (\citealt{Eggleton83})
\be
R_{\rm L} \simeq 0.462 a \left(\frac{M_{\star}}{M_{\bullet}}\right)^{1/3}.
\ee
This becomes equal to the radius of the star $R_{\star}$ below a critical semi-major axis
\begin{eqnarray}
a_{0} &\simeq& 2.17 R_{\star} \left(\frac{M_{\bullet}}{M_{\star}}\right)^{1/3} \approx 2.16\,{\rm AU}\,R_{\star,\odot}M_{\bullet,7}^{1/3}M_{\star,\odot}^{-1/3} \nonumber \\
&\approx& 2.16\,{\rm AU}\,M_{\bullet,7}^{1/3}M_{\star,\odot}^{7/15},
\label{eq:a0}
\end{eqnarray}
where in the second line we have used a mass-radius relationship $R_{\star} \simeq R_{\odot} M_{\star,\odot}^{4/5}$ appropriate for low mass main sequence stars \citep{Kippenhahn&Weigert90}. Below a critical black hole mass $M_{\bullet} \lesssim 7\times 10^{7}M_{\star,\odot}^{7/10}M_{\odot}$, the star overflows its Roche lobe outside the innermost stable circular orbit (ISCO) $R_{\rm isco} = 6GM_{\bullet}/c^{2}$ for a Schwarzschild black hole.  

The GW inspiral time at Roche contact is
\be
\tau_{\rm GW,0} \equiv \tau_{\rm GW}(a_{0}) \approx 2.8\times 10^{5}\,{\rm yr}\,M_{\star,\odot}^{13/15}M_{\bullet,7}^{-2/3}.
\label{eq:tauGW0}
\ee
Following Roche overflow, the star loses mass, primarily through the inner L1 Lagrange point, at a characteristic rate,
\be
\dot{M_{\star}} \simeq -\frac{M_{\star}}{\tau_{\rm GW}}
\ee
As the star loses mass, its radius changes according to $R_{\star} \propto M_{\star}^{p}$, where the value of $p$ depends on the properties of the star and its response to mass loss (see below).  

Combining the above results, one finds
\be
\frac{\dot{M}_{\star}}{M_{\star}} = \frac{-1}{\tau_{\rm GW,0}}\left(\frac{M_{\star}}{M_{\star,0}}\right)^{\frac{(7-12p)}{3}} = \frac{-1}{\tau_{\rm GW,0}}\left(\frac{a}{a_{0}}\right)^{\frac{(7-12p)}{(3p-1)}},
\ee
where $M_{\star,0}$ is the initial mass of the star.  For a fixed value of $p$, this results in the following evolution as a function of time $t$ after the onset of Roche-lobe overflow,
\be
\frac{M_{\star}}{M_{\star,0}} = \left(1 - \frac{12p-7}{3}\frac{t}{\tau_{\rm GW,0}}\right)^{\frac{3}{12p-7}}
\label{eq:Mevo},
\ee
%\be
%\frac{R_{\star}}{R_{\star,0}} = \left(1+ \frac{12p-7}{3}\frac{t}{\tau_{\rm GW,0}}\right)^{\frac{3p}{12p-7}}
%\ee
\be
\frac{a}{a_0} = \left(1 - \frac{12p-7}{3}\frac{t}{\tau_{\rm GW,0}}\right)^{\frac{3p-1}{12p-7}}.
\label{eq:aevo}
\ee

In reality, the value of $p$ evolves in time as the star loses mass.  Following \citet{Linial&Sari17}, for stars of initial mass $M_{\star,0} \lesssim 7.0M_{\bullet,7}^{0.25} M_{\odot}$, we have the following evolution
\be
p = \left\{
\begin{array}{lr} 0.8, &
M_{\star} > M_{\rm ad} \\
4/15, &
1.2M_{\odot}M_{\bullet,7}^{0.1} < M_{\star} < M_{\rm ad}\\
13/21, &
0.08M_{\odot} (R_{\star}/0.1R_{\odot})^{3} < M_{\star} < 1.2M_{\odot}M_{\bullet,7}^{0.1}, \\
\approx 0, &
M_{\star} < 0.08M_{\odot}(R_{\star}/0.1R_{\odot})^{3} \\
\end{array}
\label{eq:p}
\right. ,
\ee
where $M_{\rm ad} = 0.18M_{\odot}(M_{\star,0}/M_{\odot})^{17/9}M_{\bullet,7}^{-2/9}$ is the critical mass at which the GW loss timescale $\tau_{\rm GW}$ equals the Kelvin-Helmholtz cooling timescale $\tau_{\rm KH}$.  Stars with masses initially above this critical mass evolve close to the main sequence ($p = 0.8$), before evolving with $\tau_{\rm GW} = \tau_{\rm KH}$ at lower masses, with $p = 4/15$ or $p = 13/21$, depending on whether the stellar envelope is radiative or convective.  In the final line of equation (\ref{eq:p}), we have taken $p \approx 0$ for stars below the hydrogen fusion limit to account for the radius being approximately independent of mass from brown dwarfs to Jupiter scale planets (e.g.~\citealt{Chabrier+09}).

The above analysis assumes that mass transfer is stable, as occurs when upon mass loss the radius of the star decreases faster than its Roche lobe radius.  Mass loss from the star feeds an accretion disk around the SMBH.  A common assumption is that the gas disk transfers most of its angular momentum back to that of the orbit $J_{\rm orb} \simeq M_{\star}(GM_{\bullet}a)^{1/2}$, which is therefore conserved\footnote{In this stability analysis we are focusing on timescales much shorter than $\tau_{\rm GW}$, and therefore neglect angular momentum loss to gravitational radiation.}.  Since 
\be \frac{R_{\rm L}}{R_{\star}} \propto aM_{\star}^{1/3-p} \underset{\rm J_{\rm orb}=const}\approx  M_{\star}^{-5/3-p},
\ee
we see that stable mass transfer requires $p > -5/3$ in the conservative case, as is satisfied for all stages of evolution given in equation (\ref{eq:p}).  Even if no angular momentum is placed back into the stellar orbit, $R_{\rm L}/R_{\star}$ still increases upon mass loss at fixed $a$ for $p > 1/3$, as will be satisfied until the period minimum is reached and $p \approx 0$.

\begin{figure*}[!t]
\includegraphics[width=0.5\textwidth]{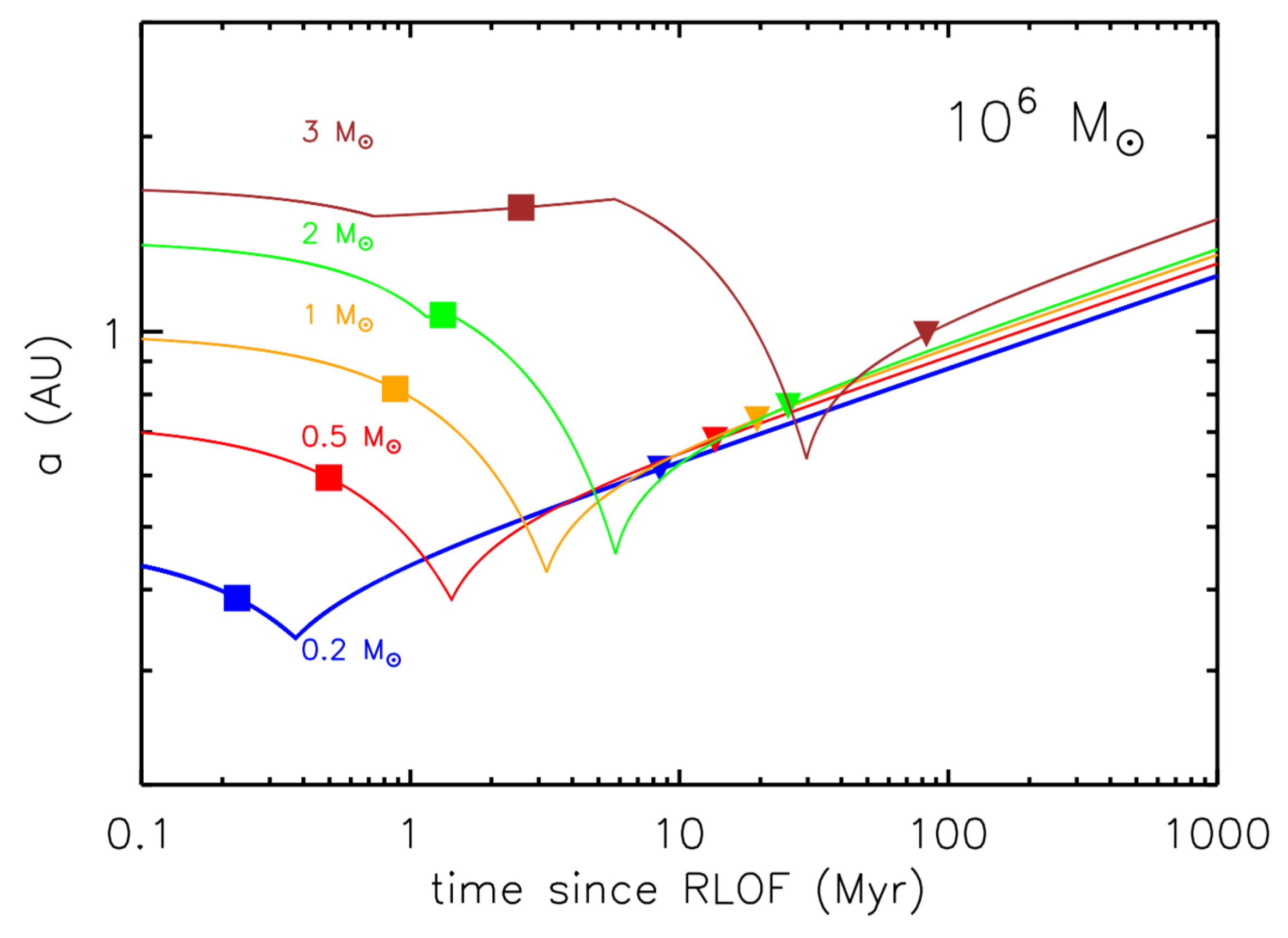}
\includegraphics[width=0.5\textwidth]{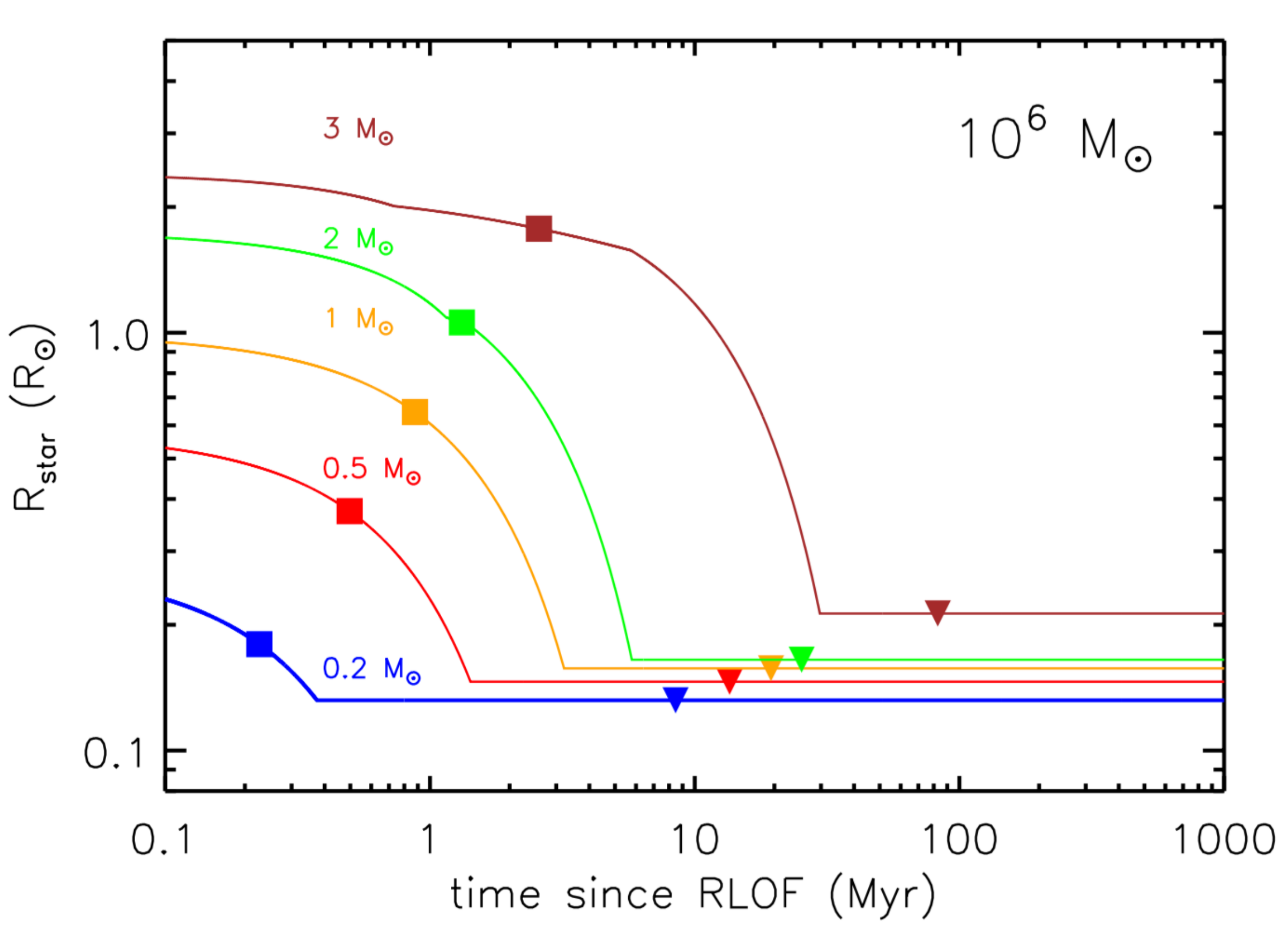}
\includegraphics[width=0.5\textwidth]{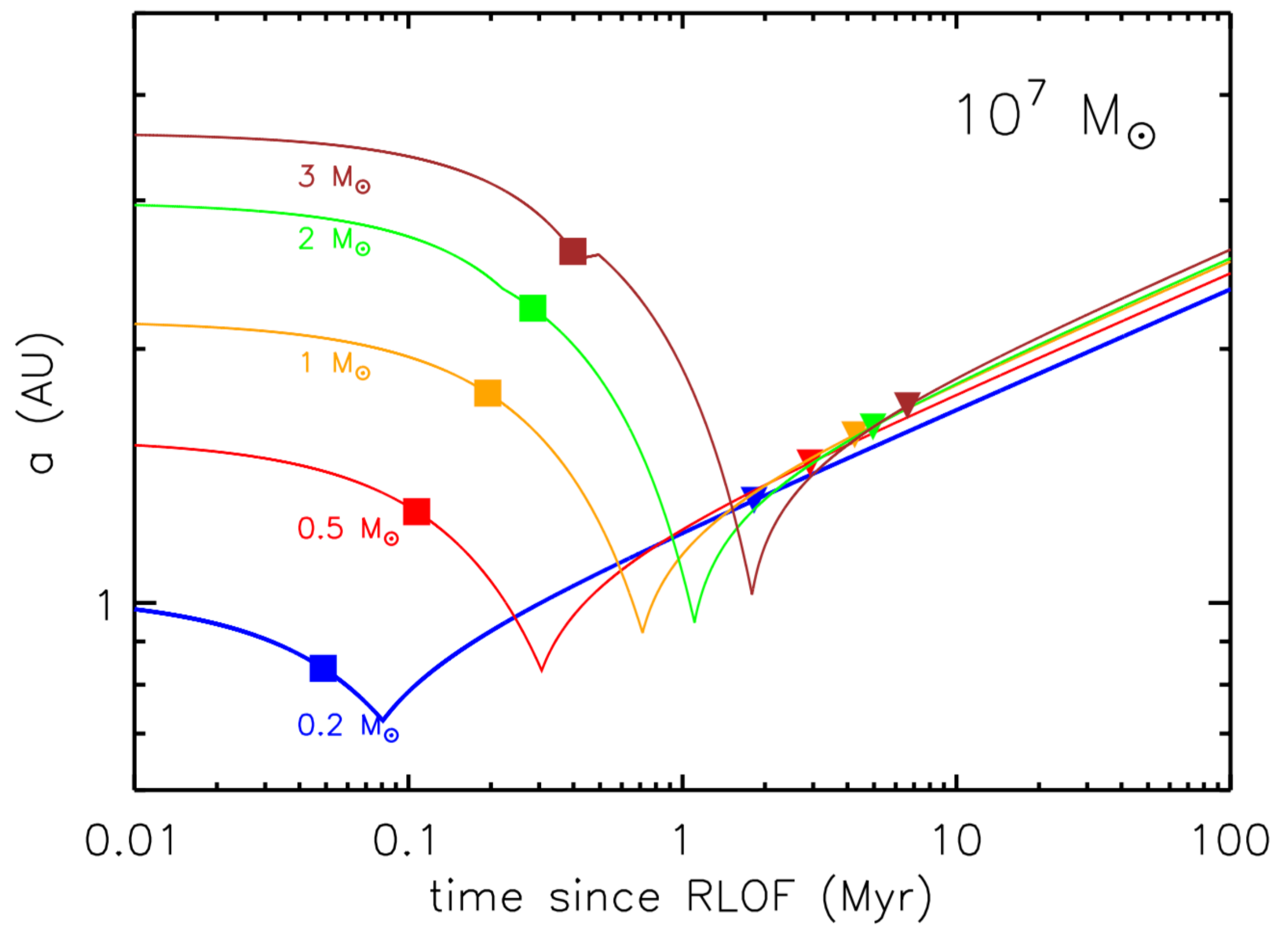}
\includegraphics[width=0.5\textwidth]{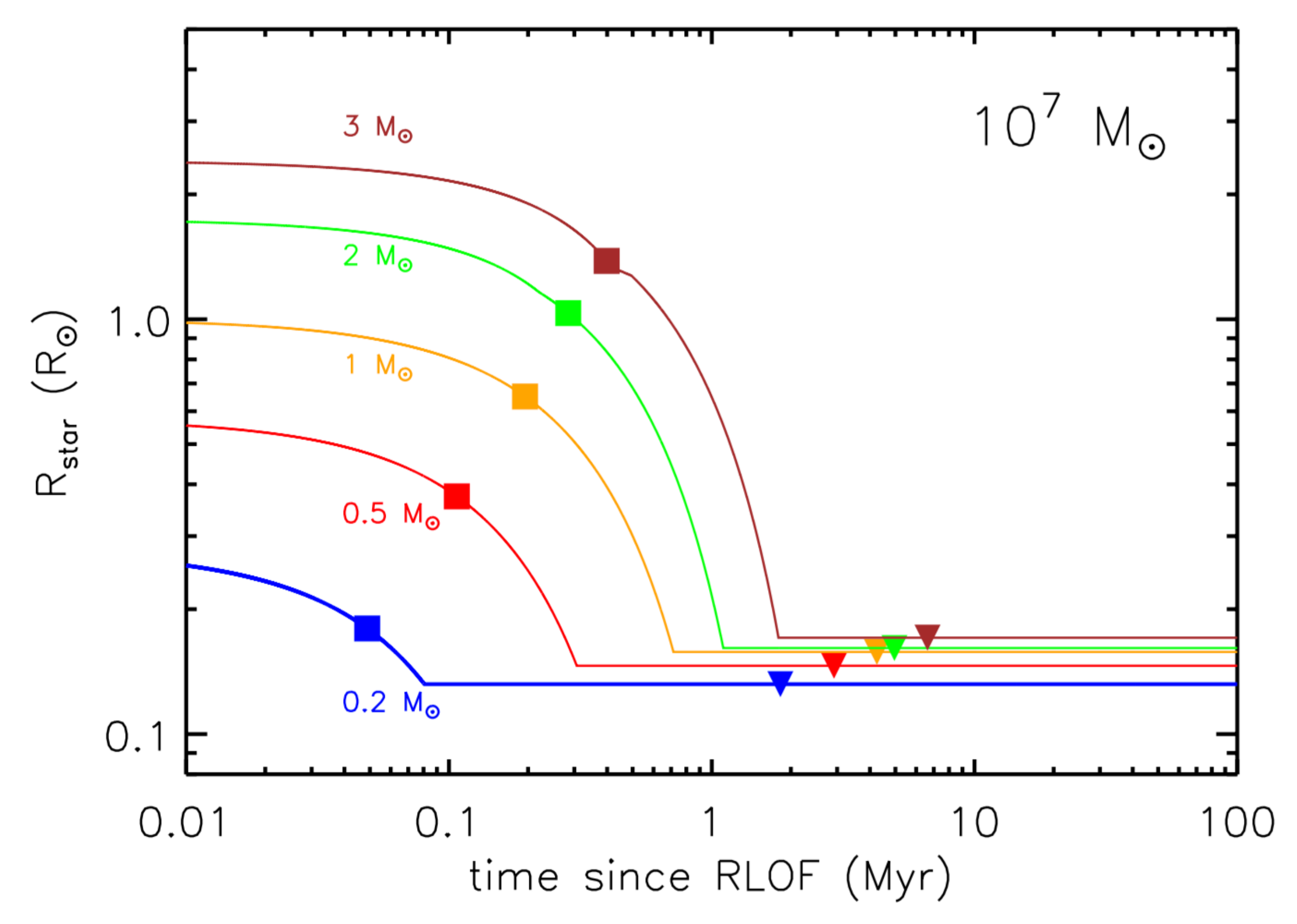}
\hspace{0.0cm}
%\includegraphics[width=0.35\textwidth]{schematic.pdf}
%\vspace{-0.4cm}
\caption{\footnotesize
{Evolution of semi-major axis (left panels) and stellar radius (right panels) as a function of time since the onset of Roche-lobe overflow onto the SMBH of mass $M_{\bullet} = 10^{6}M_{\odot}$ (top panel) and $10^{7}M_{\odot}$ (bottom panel), shown for stars of several initial masses as marked.  Squares and triangles denote, respectively, the point at which the stellar mass has decreased to one half and one tenth of its initial value.}
}
\label{fig:orbit}
\end{figure*}

Figure \ref{fig:orbit} shows the evolution of the semi-major axis and stellar radius after RLOF contact as a function of time for SMBHs of mass $M_{\bullet} = 10^{6}M_{\odot}$ (top panel) and $10^{7}M_{\odot}$ (bottom panel).  Squares and triangles denote, respectively, the point at which the stellar mass has decreased to one half and one tenth of its initial value.  

Low mass stars $0.2-0.5M_{\odot}$ lose half their mass in $0.1-0.5$ Myr, while higher mass $2-3M_{\odot}$ stars require $1-3$ Myr to do the same.  These timescales decrease with increasing SMBH mass, approximately as $\propto \tau_{\rm GW,0} \propto M_{\bullet}^{-2/3}$ (eq.~\ref{eq:tauGW0}).  
Throughout most of their inspiral evolution, stars evolve from large to small semi-major axis in the convective regime $p = 13/21$, and their semi-major axis evolves on a characteristic timescale (eqs.~\ref{eq:Mevo}, \ref{eq:aevo})
\be
\tau_{\rm GW}^{\rm RLOF} = \left(\frac{a}{|\dot{a}|}\right)_{\rm RLOF} = \chi \tau_{\rm GW} \underset{p = 13/21}= \frac{7}{2}\tau_{\rm GW}.
\label{eq:tauGWRLOF}
\ee 
This timescale is larger by a factor of $\chi = 3/(3p-1)$ than the inspiral time $\tau_{\rm GW}$ which occurs prior to the onset of mass transfer (eq.~\ref{eq:tauGW0}).

%given by equations (\ref{eq:Mevo}, \ref{eq:aevo}),
%\be \tau_{\rm GW}^{\rm RLOF} \equiv \left.\frac{a}{|\dot{a}|}\right|_{\rm RLOF} = \frac{3}{|12p-7|}\tau_{\rm GW,0} \underset{p = 13/21} = 7 \tau_{\rm GW,0} 
%\label{eq:taupark},
%\ee 
%where $\tau_{\rm GW,0}$ is the naive GW inspiral time (eq.~\ref{eq:tauGW0}).  

At late times following the period minimum ($p = 0$), the orbit evolves to larger radii as $a \propto t^{1/7}$ while the mass decreases as $M_{\star} \propto t^{-3/7}$.  This relatively slow rate of outspiral implies that the star (now technically a brown dwarf) will retain substantial mass of $\sim 0.01-0.1M_{\odot}$ for a period of time $\sim 1-10$ Myr.  As we discuss below, even such a low mass (but high density) object is sufficient to damage or disrupt a more massive star in a collision given the enormous orbital velocities on these radial scales.    
%The large prefactor predicts a substantially slower evolution once stars come into Roche contact than during their previous detached GW inspiral, which occurs on the shorter timescale of $\approx \tau_{\rm GW,0}/4$.  
%Note that throughout the entire mass transfer evolution, stellar radii vary only from $\sim$ few $R_{\odot}$ for high mass stars to $\sim 0.1R_{\odot}$ during the brown dwarf/planet outspiral phase.  

\section{Collision Between Successive EMRIs}
\label{sec:collision}

Main sequence EMRIs which undergo stable mass transfer are estimated to occur in galactic nuclei at the rate $\mathcal{R}_{\rm emri} \gtrsim 10^{-7}$ yr$^{-1}$ (\citealt{AmaroSeoane+12}).  The timescale between consecutive EMRIs of $\sim 1/\mathcal{R}_{\rm emri} \lesssim 10$ Myr is therefore comparable to the slow outspiral timescale of a star undergoing RLOF (Fig.~\ref{fig:orbit}).  This implies the probable existence at any time in a galactic nucleus of a $\sim 0.1M_{\odot}$ star or brown dwarf undergoing mass transfer evolution (e.g.~\citealt{Linial&Sari17}).  It also raises the possibility for a strong interaction or collision between successive inspiraling/outspiraling EMRIs. 

Based on the semi-major axis evolution (Fig.~\ref{fig:orbit}) and the initial mass function of stars, one common way a collision could occur is between a star of initial mass $\sim M_{\odot}$ which has already already transferred most of its mass and is now migrating outwards as a brown dwarf of mass $M_{1} \lesssim 0.1M_{\odot}$ and radius $R_{1} \approx 0.1R_{\odot}$, and another star of similar initial mass $M_{2} \sim M_{\odot} > M_1$ and radius $R_{2} \sim R_{\odot} > R_1$ which has also begun RLOF but is still moving inwards and thus would cross the orbit of $M_1$ at a distance of $a \sim 1$ AU with most of its initial mass still intact.  We consider this example as a fiducial case.  Note that, at the point of a collision, each star fills its Roche lobe and shares approximately the same semi-major axis; therefore the mean densities of the stars when they are interacting are equal, i.e. $M_{1}/R_{1}^{3} \simeq M_{2}/R_{2}^{3}.$

Our analysis in $\S\ref{sec:emri}$ assumes conservative mass transfer, which does not account for self-interaction between two EMRIs undergoing Roche-lobe overflow.  For instance, if the accretion disk of the more massive inspiraling star interferes with the ability of the less massive star's disk to feed its angular momentum back into the orbit, this could destabilize the orbit of the outspiraling brown dwarf (because $p < 1/3$ during the outspiral phase).  It is well beyond the scope of this paper to address the complex interplay between mass transfer in three body systems, and so we leave this issue to future work.  However, we note that if the EMRI rate is higher than we have assumed (e.g.~$\gtrsim 10^{-5}-10^{-6}$ yr$^{-1}$ due to the influence of massive perturbers; \citealt{Perets+07}) then even a collision between two consecutive stars which are still inspiraling ($p > 1/3$) may occur as the more massive star overtakes a less massive one.  Although the radial migration rate of the stars would differ in this case than the precise evolution predicted in $\S\ref{sec:emri}$, the qualitative collisional interaction we describe hereafter would not be altered.

Given the high rate of EMRIs of stellar mass black holes compared to those of main sequence ones, direct interactions between between inspiraling black holes and stars could be more common than star-star collisions.  However, we show in Appendix \ref{sec:A} that the collision velocity is so high that the tidal or accretion interaction between the star and the black hole passing through it is probably too small to influence the evolution of the star appreciably.  

\subsection{Conditions for a Collision}

Neither the orientation of the orbital plane or the orbital phases of the two stars will in general be aligned as they approach each other.  However, a physical collision\footnote{The orbital velocities of the stars greatly exceed their surface escape speeds, so that the effect of gravitational focusing on their cross section is negligible.} is still possible once the semi-major axes of the stars cross near a value $\sim 1$  AU (eq.~\ref{eq:a0}), as illustrated in Fig.~\ref{fig:cartoon}.  

We may neglect the comparatively slow outward radial motion of $M_1$ compared to the faster inspiral of $M_2$: the less massive $M_{1}$ can viewed as radially stationary for purposes of their interaction.  The more massive star $M_{2}$ migrates inward a distance $\delta r$ on a timescale given by $(\delta r/a)\tau_{\rm GW}^{\rm RLOF}$, where $\tau_{\rm GW}^{\rm RLOF} = \chi\tau_{\rm GW}$ is the characteristic inspiral time assuming that $M_{2}$ is undergoing mass transfer (eq.~\ref{eq:tauGWRLOF}).  The number of orbits of period $\tau_{\rm orb} = 2\pi (a^{3}/GM_{\bullet})^{1/2}$ required for $M_2$ to migrate radially by $\delta r$ is therefore
\begin{eqnarray}
&&N_{\rm GW} = \frac{\tau_{\rm GW}^{\rm RLOF}}{\tau_{\rm orb}}\frac{|\delta r|}{a} \nonumber \\
 &\approx& 1.3\times 10^{6}\chi_{3.5}\frac{a_{\rm AU}^{3/2}}{M_{\bullet,7}^{3/2}}\frac{R_{2,\odot}}{M_{2,\odot}}\left(\frac{\delta r}{2R_{2}}\right),
\end{eqnarray}
where $\chi_{\rm 3.5} \equiv \chi/3.5$.

Stars on orbits with precisely the same semi-major axis would possess identical orbital periods, modulo tiny reduced mass differences $\sim \mathcal{O}(M_{\star}/M_{\bullet}) \lesssim 10^{-6}$, and thus cannot collide.  However, stars on orbits with a small but finite semi-major axis difference $|\delta a| \lesssim R_{2}$ are still able to produce a physical collision, as their periods differ by an amount $|\delta \tau_{\rm orb}|/\tau_{\rm orb} \simeq (3/2)|\delta a|/a$ and thus will share the same orbital phase, on average, after $N_{\phi} \approx \tau_{\rm orb}/(2\delta \tau_{\rm orb}) \approx a/(3|\delta a|)$ orbits.  

A physical collision requires not only that both stars cross the same orbital phase, but also that their orbital planes cross at this phase, i.e. that they reside within a distance $l_{\perp} \sim 2^{3/2}(R_{2}b)^{1/2}$ of the line of ascending nodes of $M_{2}$ relative to the orbital plane of $M_{1}$, where $l_{\perp}$ is the approximate length parallel to the surface of $M_{1}$ at times when the radial depth of $M_{1}$ into the atmosphere of $M_{2}$ is $b \ll R_{2}$ ($b$ is also the impact parameter of the collision; see Fig.~\ref{fig:cartoon}).  This coincidence will occur only a fraction $2l_{\perp}/(2\pi a)$ of the times the phases cross, where the factor of 2 accounts for the two locations where the orbital planes cross.

Combining results, the number of orbits required for a physical collision - once their separation $|\delta a| \lesssim R_2$ is sufficiently small to enable one - is given by
\begin{eqnarray}
N_{\rm coll} &\approx& \frac{\pi}{2^{3/2}}\frac{a}{(R_{2}b)^{1/2}}N_{\phi}  \nonumber \\
&\approx& 1.7\times 10^{4} \,\frac{a_{\rm AU}^{2}}{R_{2,\odot}^{2}}\left(\frac{|\delta a|}{R_{2}}\right)^{-1}\left(\frac{b}{R_{2}}\right)^{-1/2}.
\label{eq:Ncoll}
\end{eqnarray}
The requisite condition for a single collision to occur before $M_2$ radially migrates past $M_1$ is that $N_{\rm GW}/N_{\rm coll} \gtrsim 1$ for $\delta r \lesssim 2 R_{2}$, $|\delta a| \sim R_{2}$, and $b \sim R_{2}$, where 
\begin{eqnarray}
\frac{N_{\rm GW}}{N_{\rm coll}} \approx 77 \frac{\chi_{3.5}}{a_{\rm AU}^{1/2}}\frac{R_{2,\odot}^{3}}{M_{\bullet,7}^{3/2}M_{2,\odot}}\left(\frac{|\delta a|}{R_{2}}\right)\left(\frac{\delta r}{2R_{2}}\right)\left(\frac{b}{R_{2}}\right)^{1/2}.
\label{eq:Nratio}
\end{eqnarray}
The fact that this ratio exceeds unity is a key result.  It shows that at least a single collision between the two stars is likely for all SMBH masses of interest $M_{\bullet} \lesssim 3\times 10^{7}M_{\odot}$.  

\subsection{Outcome of Stellar Collision}

\begin{figure}[!t]
\includegraphics[width=0.5\textwidth]{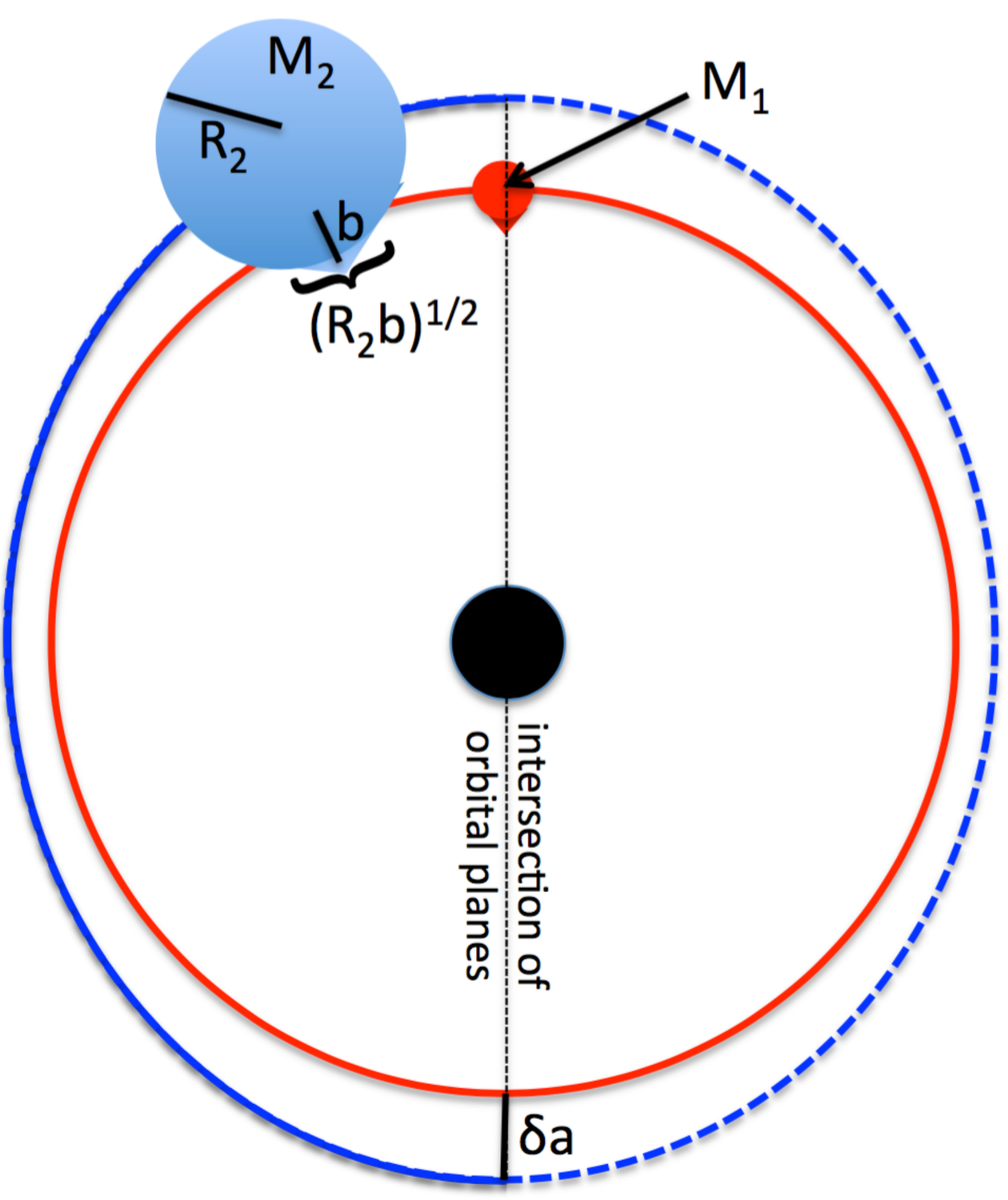}
\hspace{0.0cm}
%\includegraphics[width=0.35\textwidth]{schematic.pdf}
%\vspace{-0.4cm}
\caption{\footnotesize
{Schematic diagram (not to scale) showing circular EMRI orbits around the SMBH of the effectively stationary low mass brown dwarf $M_1$ (red orbit) and the inward-migrating star $M_2 > M_1$ (blue orbit).  The diagram is looking down on the orbital plane of $M_1$, near the time when the semi-major axes of the stars differ by an amount $\delta a \lesssim R_{2}$ and their orbital phases match along the line of ascending nodes of $M_{2}$ (defined by the plane of $M_{1}$).  When a collision occurs, $M_1$ will penetrate inside $M_2$ with an impact parameter $b$ along a chord measuring approximately $2^{3/2}(R_{2}b)^{1/2}$.  For clarity, both stars are drawn here occupying approximately the same orbital plane (though the orbit of $M_2$ is slightly inclined and flattened in projection), but in general the orbital inclinations will be misaligned and hence will only overlap along the line of ascending nodes of $M_{2}$ relative to the osculating reference plane defined by the orbit of $M_{1}$.  The orbital planes will also be evolving significantly due to Lense-Thirring precession if the SMBH is spinning.}
}
\label{fig:cartoon}
\end{figure}

We define the collision impact parameter $b$ as the distance measured from the center of $M_{1}$ to the outer edge of $M_{2}$ (Fig.~\ref{fig:cartoon}).  The impact parameter of the first collision $b = b_{\rm 1st}$ will on average be equal to half the inward radial distance $\delta r$ traveled between collisions.  This value $b_{\rm 1st} = \delta r/2$ is found by equating $N_{\rm GW} = N_{\rm coll}$ using equation (\ref{eq:Nratio}) and assuming a grazing encounter $\delta a \approx R_{2}$, $b_{\rm 1st} \ll R_{2}$.  This gives a value
\be 
\frac{b_{\rm 1st}}{R_{2}} \approx 0.06 \chi_{3.5}^{-2/3}a_{\rm AU}^{1/3}M_{\bullet,7}\frac{M_{2,\odot}^{2/3}}{R_{2,\odot}^{2}},
\label{eq:impact}
\ee
which is typically a few percent of the radius of $R_{2}$.  In most cases of interest $b_{\rm 1st} \lesssim R_{1} \sim 0.1R_{\odot}$ and hence $M_1$ will just graze the surface layers of $M_{2}$ instead of punching through its envelope at greater depth.
%By contrast, if $b_{\rm 1st} \gtrsim R_{2}$ then $M_1$ would instead punch through the low density outer envelope of $M_2$.

The orbital velocity at the time of collision is a mildly relativistic\footnote{As noted before, in this work we treat the orbital dynamics of EMRI collisions in a primarily Newtonian way, which is accurate to leading order aside from the issue of nodal precession - which we account for in the post-Newtonian approximation.  However, circular orbit speeds $v_{\rm k}\sim 0.1c$ may motivate future, fully general relativistic treatments of this scenario.}
\begin{eqnarray}
v_{\rm k} &\simeq& \left(\frac{GM_{\bullet}}{a}\right)^{1/2} \approx 9.4\times 10^{9}\,{\rm cm\,s^{-1}}M_{\bullet,7}^{1/2}a_{\rm AU}^{-1/2}. 
\label{eq:vc}
\end{eqnarray}
Depending on the mutual inclination angle $i$ of the orbital planes of $M_2$ with respect to that of $M_{1}$, the collision will occur at a relative speed 
\be
v_{\rm c} = \sqrt{2}(1-\cos i)^{1/2}v_{\rm k},
\ee
This value ranges from $v_{\rm c} = 2 v_{\rm k}$ for a head-on collision to $v_{\rm c} \ll v_{\rm k}$ for a tail-on collision.  Assuming an isotropic distribution of inclination angles, we find an average value of $\langle v_{\rm c}\rangle = (4/3)v_{\rm k}$.

\begin{figure}[!t]
\includegraphics[width=0.5\textwidth]{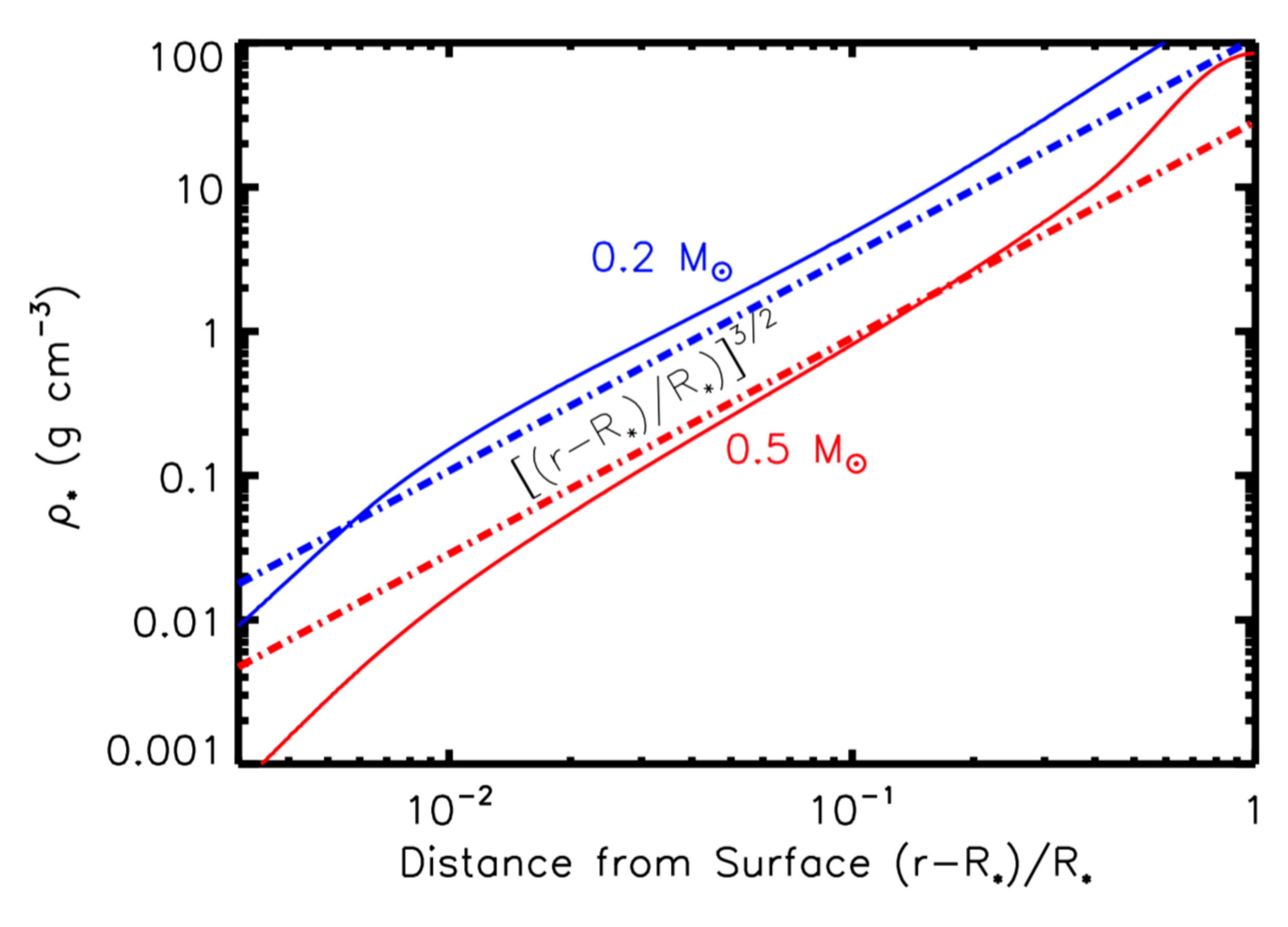}
\hspace{0.0cm}
%\includegraphics[width=0.35\textwidth]{schematic.pdf}
%\vspace{-0.4cm}
\caption{\footnotesize
{Density $\rho_{\star}$ as a function of the distance below the stellar surface $b = r-R_{\star}$ for main sequence stars of mass 0.2$M_{\odot}$ (solid red line) and $0.5M_{\odot}$ (solid blue line), compared to our adopted approximation $\rho_{\star} \simeq 0.8(M_{\star}/R_{\star}^{3})(b/R)^{3/2}$ (dashed lines).}
}
\label{fig:density}
\end{figure}

\begin{figure}[!t]
\includegraphics[width=0.5\textwidth]{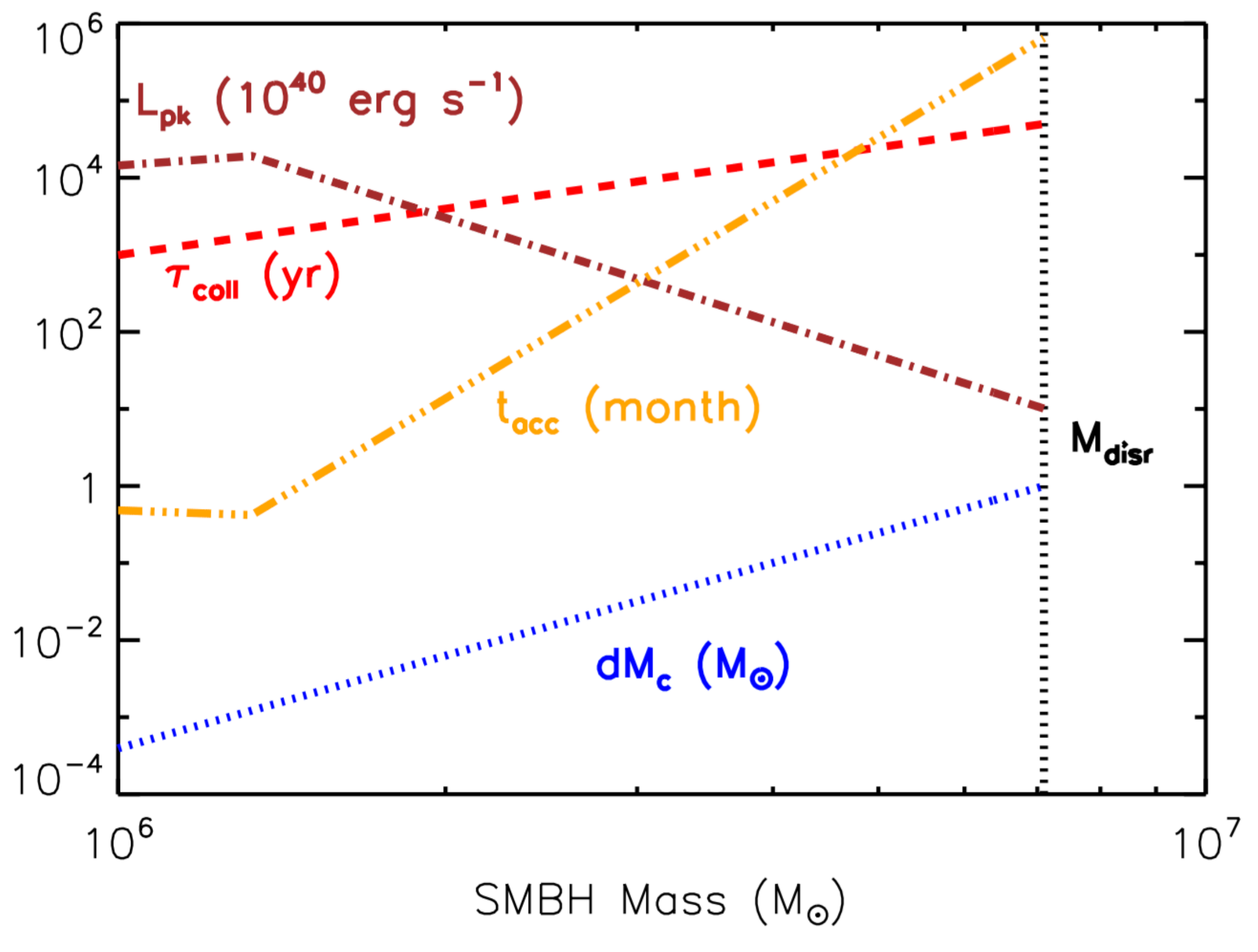}
\hspace{0.0cm}
%\includegraphics[width=0.35\textwidth]{schematic.pdf}
%\vspace{-0.4cm}
\caption{\footnotesize
{Analytic estimate of average interval between stellar collisions $\tau_{\rm coll}$ in years (dashed red line; eq.~\ref{eq:taucoll}), average mass released per collision $dM_{\rm c} = \delta M_{\rm c,max}$ in solar masses (dotted blue line; eq.~\ref{eq:dMc}) as a function of SMBH mass $M_{\bullet}$.  We have adopted characteristic parameters for the semi-major axis of the stars $a = 1$ AU, the stellar radius $R_{2} = R_{\odot}$ and mass $M_{2} = M_{\odot}$, and average collision velocity $v_{\rm c} = \langle v_{\rm k}\rangle$.  For $M_{\bullet} \gtrsim M_{\rm disr}$ the stars are almost completely disrupted in a single collision.  If the collisional mass loss is smaller than we have assumed ($\delta M_{\rm c} < \delta M_{\rm c,max}$), then the value of $M_{\rm disr}$ will increase and $\tau_{\rm coll}$ will decrease.  Also shown are the estimated peak accretion luminosity $L_{\rm pk}$ (brown dot-dashed line; eq.~\ref{eq:Lpk}) and accretion timescale $t_{\rm acc}$ (orange triple dot-dashed line; eq.~\ref{eq:tacc}) for fiducial values of the disk viscosity $\alpha = 0.01$, radius $r_{\rm g} = 1$ AU, and SMBH accretion efficiency $\eta = 0.1$. }
}
\label{fig:collision}
\end{figure}

Tidal forces between the passing stars impart modest accelerations because the collision velocity is typically two to three orders of magnitude higher than the surface escape speeds of the stars (see Appendix \ref{sec:A} for further discussion).  However, the direct, albeit grazing, physical collision between the surfaces of the stars will result in powerful shock heating, leading to gaseous mass loss.  Quantifying the collision properties, such as the total ejecta mass and its dependence on the impact parameter $b/R_{2}$, would require 3D hydrodynamical simulations well beyond the scope of this work.  Numerical simulations of stellar collisions exist in the literature (e.g.~\citealt{Freitag&Benz05}), however these are generally for much lower velocities than considered here and do not involve stars which already fill their Roche surfaces.  In what follows, we instead provide a rough analytic estimate of the mass loss and its impact on the subsequent evolution of the stars.  This crude treatment is justified in part because our qualitative conclusions are insensitive to the precise amount of mass loss per collision.

We approximate $M_2$ as a polytrope of index $n = 3/2$ appropriate for a lower main sequence convective star, in which case its density at a depth $b$ from its outer edge is very approximately given by
\be \rho_{\star} \approx 0.8\left(\frac{M_{2}}{R_{2}^{3}}\right)\left(\frac{b}{R_{2}}\right)^{3/2}, \ee  
as is shown in Fig. \ref{fig:density}.  A larger value of $n = 3$ would be appropriate for the outer layers of stars with radiative envelopes (such as main sequence stars more massive than the Sun), but the subsequent treatment is easily generalized.  We neglect non-spherical distortion of the star caused by its Roche-lobe filling shape.

The drag force on $M_1$ as it grazes the atmosphere of $M_2$ is approximately $F_{\rm d} \approx A(\rho_{\star}v_{\rm c}^{2}/2)$, where $A \approx 2R_{1} b $ is the effective cross section of the encounter, assuming $ b \lesssim R_{1}$ (Fig.~\ref{fig:cartoon}).  The energy dissipated by shock heating as $M_1$ passes across the distance $l_{\perp} \approx 2^{3/2}(bR_{2})^{1/2}$ through $M_2$ is therefore 
\begin{eqnarray}
E_{\rm c} &\approx& F_{\rm d}l_{\perp} \approx 2.3M_{2}v_{\rm c}^{2}\frac{R_{1}}{R_{2}}\left(\frac{b}{R_{2}}\right)^{3}  \\
&\approx& 7.1\times 10^{46}\,{\rm erg}\,\,\,M_{\bullet,7}\frac{M_{2,\odot}}{a_{\rm AU}}\left(\frac{v_{\rm c}}{\langle v_{\rm c}\rangle}\right)^{2}\left(\frac{b}{0.01R_{2}}\right)^{3},\nonumber
 \label{eq:Ec}
\end{eqnarray}
where in the second line and hereafter we take $R_{1} = 0.1R_{2}$.  This energy is usefully compared to the gravitational binding energy of the stars
\be
E_{\rm b} \approx \frac{GM_{\star}^{2}}{R_{\star}} \approx 4\times 10^{48}\,{\rm erg}\, \left(\frac{M_{\star}}{M_{\odot}}\right)^{2}\left(\frac{R_{\star}}{R_{\odot}}\right)^{-1}.
\label{eq:Eb}
\ee
 The ratio of the collision heating energy $E_{\rm c}$ to the binding energy $E_{\rm b,2}$ of $M_2$ is given by
\begin{eqnarray}
 \frac{E_{\rm c}}{E_{\rm b,2}} &=& 0.02
\frac{M_{\bullet,7}}{a_{\rm AU}}\frac{R_{2,\odot}}{M_{2,\odot}}\left(\frac{v_{\rm c}}{\langle v_{\rm c}\rangle}\right)^{2}\left(\frac{b}{0.01R_{2}}\right)^{3} \nonumber \\
&\underset{b = b_{\rm 1st}}\approx& 3.8 \chi_{3.5}^{-2}
M_{\bullet,7}^{4}R_{2,\odot}^{-5}M_{2,\odot}\left(\frac{v_{\rm c}}{\langle v_{\rm c}\rangle}\right)^{2},
\label{eq:Mmax}
\end{eqnarray}
where in the second line we have used the characteristic impact parameter $b_{\rm 1st}$ (eq.~\ref{eq:impact}) of the first collision.

We expect complete disruption of the star if $E_{\rm c} \gg E_{\rm b,2}$.  This occurs for collisions with impact parameter well above a critical value $b_{\rm disr}$ given by
\be
\frac{b_{\rm disr}}{R_{2}} \approx 0.039\left(\frac{a_{\rm AU}}{M_{\bullet,7}}\frac{M_{2,\odot}}{R_{2,\odot}}\right)^{1/3}\left(\frac{v_{\rm c}}{\langle v_{\rm c}\rangle}\right)^{-2/3},
\label{eq:bdisrupt}
\ee
where we have taken $\chi = 3.5$.  This condition is achieved in the first collision ($b_{\rm disr} = b_{\rm 1st}$; eq.~\ref{eq:impact}) if the SMBH mass greatly exceeds a critical value
\be
M_{\bullet,\rm disr} = 7.1\times 10^{6}M_{\odot} R_{\rm 2,\odot}^{5/4}M_{2,\odot}^{-1/4}\left(\frac{v_{\rm c}}{\langle v_{\rm c}\rangle}\right)^{-1/2}
\label{eq:Mdisr}
\ee
For $M_{\bullet} \lesssim M_{\bullet,\rm disr}$ we instead have $E_{\rm c} \ll E_{\rm b,2}$ and both stars will instead survive the first collision at least partially intact.  Although the above calculation is quite approximate and should be refined by future hydrodynamical simulations, the steep scaling $E_{\rm c}/E_{\rm b, 2} \propto M_\bullet^4$ indicates two clear regimes for EMRI collisions: around larger SMBHs, a first encounter is likely destructive, while many encounters can occur around smaller SMBHs.

Any collision will result in some mass loss from both stars.  We assume that most of the total mass loss originates from the more massive star $M_2$, motivated as follows.  As $M_{1}$ passes through the outer layers of $M_2$, the ram pressure of the interaction will drive dual shocks, through the outer layers of both stars.  However, the density $\rho_{2}$ of $M_{2}$ at the collision depth is typically 2 orders of magnitude or more lower than the density $\rho_{1}$ of $M_{1}$ (both stars have equal {\it mean} densities, but the collision occurs comparatively closer to the surface of $M_{1}$ than $M_{2}$).  Since $\rho_{1} \gg \rho_{2}$ the shock through $M_{1}$ will move with a velocity larger by a factor of ($\rho_{2}/\rho_{1})^{1/2}$ than the shock through $M_{2}$ and thus will carry an energy flux larger by the same factor of $\gtrsim 10$ (which generally is comparable to or exceeds the ratio of the gravitational binding energy of the two stars).

In what follows, we make the crude approximation that the fractional mass lost from $M_{2}$ by the collision, $\delta M_{\rm c}$, is equal to the ratio of shock-deposited energy to the gravitational binding energy.  In other words, we take
\be \frac{\delta M_{\rm c}}{M_2} = \frac{\delta M_{\rm c,max}}{M_2} \approx  \frac{E_{\rm c}}{E_{\rm b,2}},
\label{eq:dMc}
\ee  
 as given in equation (\ref{eq:Mmax}).  Equivalently,
\be \delta M_{\rm c,max} \approx 3.93M_{\odot} \chi_{3.5}^{-2}
M_{\bullet,7}^{4}R_{2,\odot}^{-5}M_{2,\odot}^{2}\left(\frac{v_{\rm c}}{\langle v_{\rm c}\rangle}\right)^{2}. 
\label{eq:Mc2} \ee
%In $\S\ref{sec:transient}$ we discuss the observational consequences of such transient bursts of mass loss.  
Eq.~\ref{eq:dMc} provides only a rough upper limit on $\delta M_{\rm c}$; the true amount of mass loss could be significantly lower if low-density outer layers of $M_2$ are ejected at speeds $\gg (GM_2/R_2)^{1/2}$.  As we discuss in $\S\ref{sec:conclusions}$, our results for the long-term evolution of the stars are not qualitatively altered if the mass loss per collision is substantially lower than this maximum (though the range of SMBH masses responsible for the most interesting behavior will shift to higher values).

\subsection{Multiple Collision Evolution}

If the stars survive their first collision ($E_{\rm c} \ll E_{\rm b,2}$, $\delta M_{\rm c} \ll M_{2}$), then it is natural to ask whether and when subsequent collisions will occur.  Will the stars collide again immediately, on the next orbit, or only after a significant delay required for the orbital phases to realign?  

The semi-major axes of $M_1$ and $M_2$ differ by $\delta a \approx R_{2}$ at the time of the first collision, and hence their orbital periods differ by a factor $\delta \tau_{\rm orb}/\tau_{\rm orb} = (3/2)(\delta a/a) \approx (3/2)(R_{2}/a)$.  The stars therefore accumulate a per-orbit phase difference of $\delta \phi/2\pi = \delta \tau_{\rm orb}/\tau_{\rm orb} = (3/2)(R_{2}/a)$, which is coincidentally comparable to the maximum phase difference between the stars that allows for a collision at the common line of ascending nodes, $\delta \phi_{\rm max} =2 R_{2}/(2\pi a)$.  Therefore, we conclude that the second collision will not generally occur on the next immediate orbit, but will instead be delayed by at least another $\sim N_{\rm coll} \gtrsim 10^{4}$ (eq.~\ref{eq:Ncoll}) orbits, corresponding to a minimum time delay of 
\be
\tau_{\rm coll}^{\rm min} = N_{\rm coll}\tau_{\rm orb} \approx 57 \,\,{\rm yr}\,\,\,\frac{a_{\rm AU}^{7/2}}{M_{\bullet,7}^{1/2}R_{2,\odot}^{2}}\left(\frac{b}{0.01R_{2}}\right)^{-1/2}.
\label{eq:taucoll1}
\ee

Even were this differential phase accumulation insufficient to prevent an immediate second collision, Lense-Thirring precession would cause the angular momenta of the stellar orbits to precess rapidly about the spin axis of the SMBH.  This will advance the line of nodes of each orbit by a large fractional angle.  At leading post-Newtonian order, and assuming circular orbits, this angular shift is (e.g., \citealt{Merritt+10})
\be
\frac{\Delta\Omega}{2\pi} = 2 \chi_{\bullet}\left(\frac{a}{R_{\rm g}}\right)^{-3/2} \approx 0.06 \chi_{\bullet} a_{\rm AU}^{-3/2}M_{\bullet,7}^{3/2},
\label{eq:LT}
\ee
where $R_{\rm g} \equiv GM_{\bullet}/c^{2}$ and $-1<\chi_{\bullet}<1$ is the dimensionless spin parameter of the SMBH.

%\footnote{Change in the orbital energy due to collisions will cause the orbits of both $M_{1}$ and $M_{2}$ to acquire mild eccentricities.  However, the magnitude of the collision energy required to induce even a small eccentricity $e \sim 0.01$ is comparable to the binding energy of the stars.}  

Over the minimum time interval before the second collision $\tau_{\rm coll}^{\rm min}$, GW radiation will move $M_2$ closer to $M_1$ on average by a distance $\delta a \approx  2b_{\rm 1st}$ (eq.~\ref{eq:impact}).  If this were the whole story, then according to equation \ref{eq:Ec}, one would expect the impact parameter of the second collision to be $\sim 3$ times higher than the first and (according to equation \ref{eq:Ec}) the energy released to be $\sim 3^{3} \sim 30$ times stronger.  Likewise, the third collision would be $\sim 30$ times stronger than the second, and so on; this sequence would rapidly terminate in a single, final, disruption once $b \gtrsim b_{\rm disr}$ (eq.~\ref{eq:bdisrupt}).  

However, this simple runaway argument neglects the impact of mass loss on the radial separation between the stars.\footnote{Change in the orbital energy due to collisions will cause the orbits of both $M_{1}$ and $M_{2}$ to acquire mild eccentricities.  However, the magnitude of the collision energy required to induce even a small eccentricity $e \sim 0.01$ is comparable to the binding energy of the stars.}  
Under adiabatic mass loss, the radius of each star will expand slightly according to $R_{\star} \propto M_{\star}^{-1/3}$ (for assumed adiabatic index $\gamma = 5/3$).  More importantly, the semi-major axis of each star will increase by a much larger amount $\delta a_{\rm c}/a \approx 2 (\delta M_{\rm c}/M_{\star}).$  This expression assumes conservative mass transfer: the gaseous disk created by the collision transfers its angular momentum back into the orbit of the mass-losing star with high efficiency.

Given our expectation that the fractional mass loss from $M_{2}$ will exceed that from $M_{1}$, the main effect on the system is to increase the semi-major axis of $M_{2}$ - and thus the radial separation of the orbits of $M_{1}$ and $M_{2}$ - by an amount
\be
 \frac{ \delta a_{\rm c}}{R_{\rm 2}} \approx 2\frac{a}{R_{2}}\frac{\delta M_{\rm c}}{M_2}.
\label{eq:bout}
\ee
Again, our assumption is that most of the angular momentum lost by $M_{2}$ is placed back into its orbit, as opposed to the orbit of $M_{1}$.  This is justified by our expectation that most of the mass loss from $M_{1}$, despite being struck obliquely by $M_{2}$, will nevertheless occur quasi-isotropically, forming a gaseous disk which lies in roughly the same orbital plane of $M_{1}$.

Since this change in separation between the stars is typically much larger than the characteristic impact parameter of the first collision $b_{\rm 1st}$ (eq.~\ref{eq:impact}), this will introduce an additional delay until the next collision beyond the minimum value (eq.~\ref{eq:taucoll1}) set by the double alignment of orbital phase and nodal line.  Specifically, the time required to traverse this distance $\delta a_{\rm c}$ through gravitational wave radiation is
\begin{eqnarray}
\tau_{\rm coll} &\approx& \tau_{\rm GW}\left(\frac{\delta a_{\rm c}}{a}\right) \approx 2\tau_{\rm GW}\left.\frac{\delta M_{\rm c}}{M_{\rm 2}}\right|_{b = b_{\rm 1st}}, \nonumber \\
&\approx & 9.9\times 10^{4}\,{\rm yr}\,\, \chi_{3.5}^{-2}\frac{a_{\rm AU}^{4}M_{\bullet,7}^{2}}{R_{2,\odot}^{5}}\left(\frac{v_{\rm c}}{\langle v_{\rm c}\rangle}\right)^{2}\left(\frac{\delta M_{\rm c}}{\delta M_{\rm c,max}}\right). \nonumber \\
\label{eq:taucoll}
\end{eqnarray}
Here we have used equations (\ref{eq:Mmax},\ref{eq:dMc}) and have calculated the GW inspiral time $\tau_{\rm GW}$ neglecting mass transfer effects (eq.~\ref{eq:tauGW}) because, for most of its inward return, $M_{2}$ is no longer overflowing its Roche Lobe (though we have retained the $\chi$ dependence on the impact parameter $b_{\rm 1st}$).  This delay could be substantially shorter than estimated here if the mass loss per collision is much less than $\delta M_{\rm c,max}$ (\ref{eq:dMc}).

After this inspiral, the characteristic impact parameter and mass loss of each such collision will be similar to the first one, $b \approx \delta b_{\rm 1st}$.  Each collision will also release a comparable amount of mass as the first (eq.~\ref{eq:Mc2}).  Neglecting slow changes to the stellar properties caused by the collisions, the star will be completely destroyed over a total number of collisions very approximately given by
\begin{eqnarray}
N_{\rm c} &\sim& \left.\frac{M_2}{\delta M_{\rm c}}\right|_{b = b_{\rm 1st}} \nonumber \\
&\approx& 0.26\chi_{3.5}^{2}
M_{\bullet,7}^{-4}R_{2,\odot}^{5}M_{2,\odot}^{-1}\left(\frac{v_{\rm c}}{\langle v_{\rm c}\rangle}\right)^{-2}\left(\frac{\delta M_{\rm c}}{\delta M_{\rm c,max}}\right)^{-1}.
\label{eq:Ncoll2}
\end{eqnarray}
This number is $\gg 1$ for low mass SMBHs and approaches unity for $M_{\bullet} = M_{\bullet, \rm disr}$ (eq.~\ref{eq:Mdisr}).  For $M_{\bullet} \ll M_{\bullet,\rm disr}$, the total duration of the interaction is just given by the gravitational wave inspiral time
\be
\tau_{\rm tot} = N_{\rm c}\tau_{\rm coll} \approx 2\tau_{\rm GW}, 
\label{eq:tautot}
\ee 
which ranges from $\sim 10^{4}-10^{6}$ yr, depending on the SMBH mass.  These different regimes of collisions are summarized in Figure \ref{fig:collision}.  

\section{Transient SMBH Accretion Events}
\label{sec:transient}

We have shown that collisional evolution between consecutive stellar EMRIs around SMBHs of mass $\lesssim 10^{7}M_{\odot}$ can result in repeated, prompt episodes of gas formation; the colliding stars liberate a mass $M_{\rm g} \le \delta M_{\rm c} \sim 10^{-4}-1M_{\odot}$, on a radial scale of $r_{\rm g} \sim 0.5-2$ AU, at periodic intervals of decades to centuries or longer, and lasting for a total duration of up to a million years.  We now describe the observable signatures of these punctuated episodes of gas production.

By equating the thermal energy released per unit volume with the specific heat of radiation-dominated matter, $a T_{\rm c}^{4} \approx \rho_{\star}v_{\rm c}^{2}/2$, the immediate temperature of the shocked stellar material is $T_{\rm c} \sim 10^{8}$ K even for a head-on collision, too low to result in nuclear burning up to $^{56}$Ni.  Without such a radioactive heating source, the expanding ejecta will cool rapidly via adiabatic expansion before becoming transparent and radiating the remaining thermal energy.  This will produce a dim rapidly-evolving transient which would itself be very challenging to detect (see also \citealt{Balberg+13}).  

A more important source of luminosity is the gravitational energy liberated by the accretion of the gaseous ejecta onto the SMBH.  The ejected gas will possess an angular momentum similar to that of the stars, resulting in the formation of circular disk on a similar radial scale of $\lesssim r_{\rm g}$.  Internal stresses (likely magneto-turbulent in origin) within the accretion flow will transport this angular momentum outwards, allowing the gas mass $M_{\rm g}$ to accrete inwards on the characteristic accretion timescale
\be
t_{\rm acc} \approx \frac{1}{\alpha}\frac{1}{\Omega_k}\left(\frac{h}{r_{\rm g}}\right)^{-2} \approx 18.4\,{\rm d}\,\,\frac{r_{\rm g,AU}^{3/2}}{\alpha_{-1}M_{\bullet,7}^{1/2}}\left(\frac{h}{0.1r_{\rm g}}\right)^{-2},
\label{eq:tpk}
\ee
where $\Omega_{k} = (GM_{\bullet}/r_{\rm g}^{3})^{1/2}$ is the orbital angular velocity of the disk, $\alpha/=0.1\alpha_{-1}$ is the Shakura-Sunyaev viscosity parameter, and $h$ is the vertical scale height of the disk.

Following deposition of the gaseous material, the accretion rate will quickly rise to a peak value of $\dot{M}_{\rm pk} \sim M_{\rm g}/t_{\rm acc}$ (\citealt{Pringle81}).  For a radiation-pressure dominated disk with electron scattering opacity $\kappa_{\rm es}$, the vertical scale-height of the disk obeys $h = 3\kappa \dot{M}/(8\pi c) = \frac{3}{2\eta}(\dot{M}/\dot{M}_{\rm Edd})(GM_{\bullet}/c^{2}) \simeq 2.2\times 10^{13}(\dot{M}/\dot{M}_{\rm Edd})M_{\bullet, 7}$ cm, where $\dot{M}_{\rm Edd} \approx 1.6\times 10^{25}M_{\bullet,7}$ g s$^{-1}$ is the Eddington accretion rate for radiative efficiency $\eta = 0.1$.  Combining results, we find that
\be
\frac{\dot{M}_{\rm pk}}{\dot{M}_{\rm Edd}} \approx 5.9\times 10^{-2}\left(\frac{M_{\rm g}}{10^{-3}M_{\odot}}\right)^{-1}M_{\bullet,7}^{-3/2}r_{\rm g,AU}^{7/2}\alpha_{-1}^{-1}
\ee
with
\be
\frac{h}{r_{\rm g}} \approx 0.087\left(\frac{M_{\rm g}}{10^{-3}M_{\odot}}\right)^{-1}M_{\bullet,7}^{-1/2}r_{\rm g,AU}^{5/2}\alpha_{-1}^{-1}
\ee
and
\be
t_{\rm acc} \approx 24\,{\rm d}\,\,\ \frac{\alpha_{-1}M_{\bullet,7}^{1/2}}{r_{\rm g,AU}^{7/2}}\left(\frac{M_{\rm g}}{10^{-3}M_{\odot}}\right)^{2},
\label{eq:tacc}
\ee
The resulting peak accretion luminosity is given by
\begin{eqnarray}
&&L_{\rm pk} = \eta \dot{M}_{\rm pk}c^{2} \nonumber \\
&\approx& 8.5\times 10^{43}\,{\rm erg\,s^{-1}}\,\frac{r_{\rm g,AU}^{7/2}}{M_{\bullet,7}^{1/2}\alpha_{-1}}\left(\frac{M_{\rm g}}{10^{-3}M_{\odot}}\right)^{-1},
\label{eq:Lpk}
\end{eqnarray}
where we have assumed a radiative efficiency $\eta = 0.1$.

For a characteristic value of $M_{\rm g} \sim 10^{-3}M_{\odot}$, as is expected in our fiducial models for $M_{\bullet} \approx 2\times 10^{6}M_{\odot}$ (eq.~\ref{eq:Mc2}), we predict accretion-powered transients of peak luminosity $\approx 10^{44}$ erg s$^{-1}$ and characteristic timescales $t_{\rm acc} \approx 40~{\rm days}$.  However, these properties scale sensitively with SMBH mass.

These precise scalings, in particular the unintuitive inverse relationship between peak accretion rate and gas disk mass, must be taken with caution.  It is well known that the thermal and viscous stability of radiation pressure dominated disks remains an open issue (e.g.~\citealt{Hirose+09,Jiang+16,Sadowski&Narayan16}), and the true behavior of these disks could differ significantly from the expectations of the $\alpha$-models.  

The evolution of the accretion rate at times after the peak ($t \gg t_{\rm acc}$) depends on the interaction between the outer edge of the gaseous disk and the stars.  If most of the angular momentum of the disk is transferred back into the stellar orbit, then - in contrast to circularized TDE disks \citep{Cannizzo+90} - the disk will not freely spread outwards beyond the orbit of the star.  In this case the accretion rate will decay exponentially on a timescale set by the initial viscous timescale $t_{\rm acc}$, i.e. we expect a bolometric light curve of the form
\be
L(t) \sim L_{\rm pk}e^{-t/t_{\rm acc}}, \,\,\,\, t \gtrsim t_{\rm acc}
\label{eq:exp}
\ee  

On the other hand, a portion of the disk may be free to viscously spread outwards beyond the orbit of the star.  This could occur if the angular momentum of the gas produced by the collision places it into an orbital plane significantly different either star.  It could also occur if the Bardeen-Petterson effect \citep{Bardeen&Petterson75} aligns the disk into a different plane from both stars on a timescale short compared to the disk evolution timescale; such misalignment could also be accomplished by differential nodal precession between stellar orbits and the disk.  In this case, the mass accretion rate would instead decay at late times as a power-law, $L \propto t^{-\alpha}$, where $\alpha \sim 1.1-1.3$ (e.g.~\citealt{Cannizzo+90,Shen&Matzner14}), although we caution that this evolution could be complicated by possible thermodynamic state changes from radiation-pressure to gas-pressure dominated regimes (\citealt{Shen&Matzner14}).  In both cases, the predicted evolution differs from the canonical $L(t) \propto t^{-5/3}$ prediction for the late-time decay of the mass fall-back rate in TDEs.

\subsection{TDE Imposters}

The flare timescales and bolometric luminosities we predict from colliding main sequence EMRIs overlap with those of observed tidal disruption flare candidates (\citealt{Komossa15}).  However, unlike TDEs, which are singular events, a colliding EMRI pair may produce {\it hundreds or thousands of mass production events} (eq.~\ref{eq:Ncoll2}).  This large number may be sufficient to at least partially overcome the $\sim 2-3$ order of magnitude deficit between the predicted TDE and circular EMRI rate, allowing flares of the collisional EMRIs described here to be detected.  This would especially be true if a large fraction of TDEs are ``dark" due to inefficiency associated with accretion at super-Eddington rates (\citealt{Strubbe&Quataert09,Metzger&Stone16}) or the process of debris circularization of the highly elliptical streams produced in most TDEs (\citealt{Guillochon&RamirezRuiz15,Dai+15,Hayasaki+16}).    

The total radiated energy inferred from optical TDE candidate flares ranges from $E_{\rm rad} \approx 10^{49}$ erg for iPTF16fnl (\citealt{Blagorodnova+17}) and $3\times 10^{50}$ erg for PS1-11af (\citealt{Chornock+14}) to values up to $\sim 10$ times higher, as compiled in Figure \ref{fig:Erad}.
These radiated energies correspond to accreted masses in the range $\sim 10^{-4}-10^{-2} M_{\odot}$ for an assumed radiative efficiency of $\eta = 0.1$.  These often low radiated energies have been described as a ``missing energy problem'' \citep{Piran+15}, though this problem has many possible resolutions in the TDE paradigm, including low accretion or radiative efficiencies in TDEs (e.g.~\citealt{Piran+15, Metzger&Stone16}), large bolometric corrections \citep{vanVelzen+16}, or the lower accreted mass of a partial tidal disruption event \citep{Chornock+14}.  

Alternatively, these low radiated energies may indicate that some of the observed TDE candidates are in fact just one in a sequence of quasi-periodic EMRI collision flares, each producing a low-mass accretion transient.  One way to distinguish these scenarios is to search for additional flares from the same galactic nucleus after the initial burst.  The required wait time will usually be too long to serve as a definitive test (up to $\sim 10^5~{\rm yr}$ for fiducial parameters; see eqs. \ref{eq:taucoll1}, \ref{eq:taucoll}), though because $\tau_{\rm coll} \propto R_{2, \odot}^{-5}$ and $\tau_{\rm coll}^{\rm min} \propto R_{2, \odot}^{-3/2}$, repeated flares can happen on timescales shorter than a decade if the victim star is a few times the mass of the Sun, or is a sub-solar mass star that has bloated due to shock heating.  A shorter flare duration would occur if the mass loss per collision is much less than our conservative upper limit on its value (eq.~\ref{eq:dMc}).

The accretion timescale of the gaseous disks produced by stellar collisions, $t_{\rm acc}$ (eq.~\ref{eq:tacc}), is  short compared to the interval between consecutive collisions.  This implies that the light curve would appear as periodic spikes in luminosity, separated by quasi-periodic intervals ranging from decades to hundreds of millenia.  Such quasi-periodic flaring may have been observed in IC3599, a Seyfert galaxy which produced a large-amplitude nuclear X-ray outburst in the 1990s (\citealt{Brandt+95,Grupe+95}), initially interpreted as a TDE.  However, IC3599 was then observed to repeat its flaring behavior \citep{Grupe+15,Campana+15}.  Based on modeling the light curves and disk temperature evolution, \citet{Campana+15} claim the existence of 3 outbursts with a separation of $\approx$ 10 years, each reaching  a luminosity of $10^{43}$ erg s$^{-1}$.  

Individual flares produced from EMRI collisions may also be distinguished from standard TDE flares by differences in the predicted late-time light curve decay.  In cases when the orbits of the surviving stars absorb angular momentum from the gaseous disk, we could expect the light curve to decay exponentially at late times (eq.~\ref{eq:exp}), a feature which may in fact be observed some TDE flares (e.g.,~NGC 247, \citealt{Feng+15}; ASASSN-14li, \citealt{Holoien+16}; iPTF16fnl, \citealt{Blagorodnova+17}).  On the other hand, if a sizable fraction of the gas from the collision is able to viscously spread beyond the orbits of the stars, we would expect a power-lower light curve decay shallower than the canonical $\propto t^{-5/3}$ fall-back rate.  Such shallow decays are inferred from the X-ray sample of TDE flares compiled by \citet{Auchettl+17}.

%Such periodic outbursts could also be explained by a string of partial disruption events of the same star on a highly elliptical orbit in the case of a normal TDE.  One way to distinguish this scenario from collisional EMRIs is the lack of strict periodicity expected in the latter case due to differences in the amount of mass loss in subsequent collisions (and thus in the interval between the next collision).

One of the most puzzling mysteries of optical TDE candidates concerns their observed color temperatures, which are roughly an order of magnitude lower than those predicted from compact disk emission models (e.g., \citealt{Gezari+12,Cenko+12,Holoien+16}), which should instead peak in the extreme UV/soft X-ray band (corresponding to the disk temperature at $\sim 10R_{\rm g}$).  One explanation for this behavior is the existence of a dense layer of gas on radial scales $\gtrsim 10-100$ AU, which reprocesses much of the accretion power to lower frequencies (\citealt{Loeb&Ulmer97, Roth+16}).  In TDE scenarios, this reprocessing material could be the result of a super-Eddington wind (\citealt{Metzger&Stone16}) or highly eccentric, inefficiently circularized debris streams produced during the disruption process (e.g., \citealt{Guillochon+14,Miller15}).  

A radially extended gaseous disk is also expected to arise naturally in the EMRI collision scenario, and could provide an alternative reprocessing screen.  As discussed above, the gaseous disk deposited by a given collision at $r_{\rm g} \sim $ AU will in some cases carry its angular momentum outwards by viscously expanding \citep{Cannizzo+90} beyond the orbits of the stars.  Since total angular momentum is roughly conserved in this process, the disk mass which remains when the outer edge of the disk reaches a radius $r$ is $M(r) \sim M_{\rm g}(r/r_{\rm g})^{1/2}$.  The timescale for the gas from a collision event to reach $\sim 10-100$ AU by viscous spreading is $\sim 10^{3}-10^{4}$ yr \citep{Cannizzo+90,Shen&Matzner14}, many orders of magnitude longer\footnote{The disk evolution slows considerably once the midplane transitions from being radiation pressure dominated to being gas pressure dominated (e.g.~\citealt{Shen&Matzner14}).} than the initial viscous timescale (flare time) and potentially comparable to the entire duration of the collisional interaction between the stars, $\tau_{\rm tot}$ (eq.~\ref{eq:tautot}). 

The accumulation of spreading disks produced by each of the hundreds to thousands of collisions which occur prior to a given typical collision could therefore place up to $\sim 0.1-0.3M_{\odot}$ over radial scales of $\gtrsim 10-100$ AU, potentially sufficient to explain the observed reprocessing \citep{Roth+16}. 
Due to Lense-Thirring precession of the stellar orbits, the collision plane will rotate between subsequent encounters around the SMBH spin axis (eq.~\ref{eq:LT}), such that each gaseous disk would be created with its angular momentum pointed in a significantly different direction.  However, differential Lense-Thirring precession will generally align each transient disk with the SMBH equatorial plane \citep{Bardeen&Petterson75}, leading to a larger scale accretion flow that is coplanar.  

Once UV/X-ray irradiation from the flare intercepts the geometrically thin disk, the resulting heating will cause it to puff up vertically, blocking a larger fraction of the light than it would have given its initial vertical thickness.  If the timescale for the collision evolution is sufficiently long, with $\tau_{\rm tot} \gg 10^{4}$ yr (eq.~\ref{eq:tautot}), then gas accretion from a TDE in the same nucleus might destroy the fossil disk ($\S\ref{sec:TDEstripping}$).  However, in this case the viscous spreading-evolution of one (or several consecutive) TDE disks could itself deposit sufficient mass on large scales to replenish the repocessing disk.  Thus, our new proposed scenario for  the reprocessing layer is largely independent of the EMRI collision scenario, and could also apply to galaxies with a high TDE rate.

%Still, symmetry imposed by the direction of the BH spin should enable the cumulative debris from all collisions to possess a net angular momentum, such that their combined viscous evolution would qualitatively resemble that of a single spreading disk.  This slowly expanding structure could would cover a large fractional solid angle as subtended by the inner disk.  It would also be rotating slowly, near the local Keplerian rotational velocity, potentially consistent with the narrow widths of the optical emission line features in TDEs (e.g.~\citealt{Arcavi+14,Saxton+16}).  

This fossil reprocessing disk could also serve as source of magnetic flux, to be swept towards the black hole by the debris from future TDEs \citep{Tchekhovskoy+14,Kelley+14}, explaining why some TDEs are able to produce 
powerful magnetically-dominated jets (\citealt{Giannios&Metzger11,Tchekhovskoy+14,Kelley+14}).

\subsection{Gas Ablation in True Tidal Disruption Events}
\label{sec:TDEstripping}

The rate of TDEs around small SMBHs is typically $\sim 10^{-4}~{\rm yr}^{-1}$, with a scatter of about an order of magnitude depending on the detailed properties of the specific galaxy \citep{Stone&Metzger16}.  Given that this exceeds our fiducial rate of circular EMRI inspirals by a factor of $\sim 10^{2}-10^{4}$, the first mass-transferring star must survive this many gas production events in a galactic nucleus in order to undergo the collisional evolution described in this paper.  This section considers whether such survival is feasible in the face of mass loss due to shock-heating from the interaction between the star and the gaseous TDE accretion disk.

The TDE results in a transient accretion event of duration $\tau_{\rm acc} \sim$ months and average mass accretion rate $\dot{M}_{\rm acc} \approx M_{\rm acc}/\tau_{\rm acc}$, where $M_{\rm acc}$ is the total accreted mass.  The star's semi-major axis distance $a$ is comparable to the tidal radius of a disrupted star, and the midplane density of the gaseous TDE disk at this location is approximately given by
\be
\rho_{\rm d} = \frac{\Sigma}{2h} = \frac{\dot{M}_{\rm acc}}{6\pi \nu h} =\frac{M_{\rm acc}}{6\pi \alpha a^{3}\Omega(h/a)^{3}\tau_{\rm acc}},
\ee
where we have used the standard relationship for a steady-state disk $\dot{M}_{\rm acc} = 3\pi \nu \Sigma$ with viscosity $\nu = \alpha (h/a)^{2}a^{2}\Omega$.

In general the orbital plane of the gaseous disk will not be aligned with that of the star, and so the star will be impacted by a ``headwind" of gas at a relative velocity $v_{\rm c}$ up to twice the orbital velocity $v_{\rm k} = a\Omega$.  The shock created by the interaction of the star with the gaseous disk will ablate mass stellar mass.  

The shock will penetrate to a depth inside the star approximately given by equality between ram pressure and the interior pressure of the star,
\be
\rho_{\rm d}v_{\rm c}^{2}/2 = P_{\star} = \bar{P}_{\star}\left(\frac{\rho_{\star}}{\bar{\rho}_{\star}}\right)^{\gamma},
\ee
where $\bar{\rho}_{\star} \equiv M_{\star}/(4\pi R_{\star}^{3}/3)$ is the mean stellar density and $\bar{P}_{\star} \simeq 0.04 GM_{\star}^{2}/R_{\star}^{4}$ for a $\gamma = 5/3$ polytrope.  The shocks thus initially penetrate to a depth where the stellar density equals
\begin{eqnarray}
&&\frac{\rho_{\star,\rm sh}}{\bar{\rho}_{\star}} \approx 1.9\left(\frac{v_{\rm c}^{2}R_{\star}}{GM_{\star}}\frac{\rho_{\rm d}}{\bar{\rho}_{\star}}\right)^{3/5} \nonumber \\
 &&\approx 1.9\left[\frac{2}{9\alpha}\left(\frac{h}{a}\right)^{-3}\frac{1}{\Omega \tau_{\rm acc}}\left(\frac{v_{\rm c}}{v_{\rm k}}\right)^{2}\left(\frac{M_{\bullet}}{M_{\star}}\right)\left(\frac{M_{\rm acc}}{M_{\star}}\right)\left(\frac{R_{\star}}{a}\right)^{4}\right]^{3/5} \nonumber \\
&&\approx 2.1\times 10^{-4}\frac{M_{\bullet,7}^{3/10}}{\alpha_{-1}^{3/5}a_{\rm AU}^{3/2}}\left(\frac{h/a}{0.3}\right)^{-9/5}\left(\frac{v_{\rm c}}{v_{\rm k}}\right)^{6/5} \nonumber \\
&& \times \left(\frac{M_{\star}}{0.1M_{\odot}}\right)^{-3/5}\left(\frac{M_{\rm acc}}{M_{\star}}\right)^{3/5}\left(\frac{R_{\star}}{0.1R_{\odot}}\right)^{12/5}\left(\frac{\tau_{\rm acc}}{{\rm month}}\right)^{-3/5}.
\label{eq:rhopenetrate}
\end{eqnarray}

While the star is within the gaseous disk, the shocks will cause mass ablation from its surface.  The maximum rate of mass ablation is approximately equal to the rate at which gas passes through the shock,
\be
\dot{M}_{\star} = -\pi R_{\star}^{2}\rho_{\star, \rm sh} v_{\rm sh},
\ee
where $\pi R_{\star}^{2}$ is the cross section of the star and $v_{\rm sh} \approx (\rho_{\rm d}/\rho_{\star, \rm sh})^{1/2}v_{\rm c}$ is the velocity of the shock bring driven into the star (\citealt{McKee&Cowie75}).    

Combining results, we find that the minimum timescale for the star to be destroyed by ablation is given by
\begin{eqnarray}
&&\frac{\tau_{\rm ablate}}{\tau_{\rm acc}} \approx \frac{1}{f_{\rm d}\tau_{\rm acc}}\frac{M_{\star}}{|\dot{M}_{\star}|} = \frac{4}{3f_{\rm d}}\frac{R_{\star}}{\tau_{\rm acc} v_{\rm c}}\left(\frac{\rho_{\star, \rm sh}}{\bar{\rho}_{\star}}\right)^{-1/2}\left(\frac{\bar{\rho}_{\star}}{\rho_{\rm d}}\right)^{1/2} \nonumber \\
&=& \frac{1}{f_{\rm d}}\left(\frac{\bar{\rho}_{\star}}{\rho_{\star, \rm sh}}\right)^{\frac{1}{2}}\left(8 \alpha \left(\frac{M_{\star}}{M_{\rm acc}}\right)\left(\frac{a}{R_{\star}}\right)\left(\frac{v_{\rm k}}{v_{\rm c}}\right)^{2}\left(\frac{h}{a}\right)^{3}\frac{1}{\tau_{\rm acc}\Omega}\right)^{\frac{1}{2}} \nonumber \\
&\approx& \frac{17}{f_{\rm d}}\left(\frac{\rho_{\star, \rm sh}}{10^{-4}\bar{\rho}_{\star}}\right)^{-1/2}\frac{\alpha_{-1}^{1/2}a_{\rm AU}^{5/4}}{M_{\bullet,7}^{1/4}}\left(\frac{M_{\rm acc}}{M_{\star}}\right)^{-1/2} \nonumber \\
&&\times \left(\frac{h/a}{0.3}\right)^{3/2}\left(\frac{v_{\rm c}}{v_{\rm k}}\right)^{-1}\left(\frac{R_{\star}}{0.1R_{\odot}}\right)^{-1/2}\left(\frac{\tau_{\rm acc}}{{\rm month}}\right)^{-1/2} \nonumber \\
&\approx& 36\frac{\alpha_{-1}^{4/5}a_{\rm AU}^{2}}{M_{\bullet,7}^{2/5}}\left(\frac{M_{\rm acc}}{M_{\star}}\right)^{-4/5}\left(\frac{h/a}{0.3}\right)^{12/5} \left(\frac{f_{\rm d}}{h/a} \right)^{-1}\label{eq:tauablate}  \\
&&\left(\frac{v_{\rm c}}{v_{\rm k}}\right)^{-8/5}\left(\frac{R_{\star}}{0.1R_{\odot}}\right)^{-17/10}\left(\frac{\tau_{\rm acc}}{{\rm month}}\right)^{-1/5}\left(\frac{M_{\star}}{0.1M_{\odot}}\right)^{3/10},\nonumber
\end{eqnarray}
where $f_{\rm d}$ is the fraction of the star's orbit spent in the midplane of the disk, and in the final line we have substitute in equation (\ref{eq:rhopenetrate}).   A stellar orbit misinclined relative to the plane of the disrupted star will have $f_{\rm in} \sim h/a$, while one close to the same orbital plane would have $f_{\rm in} \sim 1$.\footnote{However, $f_{\rm in}$ may change even during the TDE itself due to Lense-Thirring precession of the stellar orbit (eq.~\ref{eq:LT}).}   

We thus see that $\tau_{\rm ablate} \gg \tau_{\rm acc}$ for fiducial parameters, and thus the star should survive at least one TDE intact.  However, the outcome is sensitive to the precise semi-major axis of the star and the total gas mass accreted in the TDE, $M_{\rm acc}$, which could be as low as $\sim 10^{-3}-10^{-2}M_{\odot} \sim 0.01-0.1M_{\star}$ (Fig.~\ref{fig:Erad}; \citealt{Metzger&Stone16}), in which case $\tau_{\rm ablate}/\tau_{\rm flare}$ would be larger by significant factor.  Also note that equation (\ref{eq:tauablate}) is likely a lower limit on the ablation timescale, because not all matter which passes through the shock will necessarily become unbound from its surface.  This is because the specific thermal energy imparted to the shocked gas $\sim v_{\rm sh}^{2} \sim (\rho_{\star, \rm sh}/\bar{\rho}_{\star})v_{\rm c}^{2} \sim 10^{-4}(GM_{\bullet}/a)$ may be comparable to the gravitational binding of the star $\sim v_{\rm esc}^{2} \sim GM_{\star}/R_{\star}$; gas stripped from the stellar surface may therefore accumulate in a wake behind the star to become re-accreted once the flare ceases. 

What is less clear from the above numbers is whether the outspiraling brown dwarf can survive the large number of expected collisions during its lifetime.  Once a second EMRI arrives on a quasi-circular orbit, the period of the collisional interaction is $\sim 10^{4-6}~{\rm yr}$ (Eq. \ref{eq:tautot}).  During this period, the expected number of TDEs is $\sim 0.1-1000$ \citep{Stone&Metzger16}.  Large portions of this range are likely surviveable for the brown dwarf.  The greater challenge for the brown dwarf is surviving long enough to greet an arriving second EMRI.  The fiducial circular EMRI rate of $\sim 10^{-7}~{\rm yr}^{-1}$ would require the brown dwarf to survive $\sim 10^{2-4}$ tidal ablation events, which will be much more challenging (Eq. \ref{eq:tauablate}).  However, we note that this fiducial circular EMRI rate could be enhanced by up to two orders of magnitude if binary orbits are perturbed outside the SMBH influence radius by massive objects (e.g. giant molecular clouds; \citealt{Perets+07}) or non-axisymmetric components of the galactic potential \citep{Hamers&Perets17}.  The large estimated galaxy-to-galaxy scatter (at fixed SMBH mass) in the classical TDE rate \citep{Stone&Metzger16} is not necessarily correlated with the comparably large scatter in the circular EMRI rate (as most of the TDEs are sourced from $\sim {\rm pc}$ scales, while tidally detached binaries can come from much larger distances), making it likely that at least a subset of galactic nuclei will produce EMRI collisions which are not seriously inhibited by ablation from interloping TDEs.

\section{Conclusions}
\label{sec:conclusions}

The inevitability of stellar collisions in galactic nuclei is well-documented (e.g.~\citealt{Ginsburg&Loeb07,Antonini+10,Balberg+13, Leigh+16a}).  However, past work has focused on singular collisions, or mergers of stars on eccentric orbits located far outside the tidal radius of the SMBH.  Given the relatively large radius of gas produced in such collisions, any resulting accretion-powered flare produced by these collisions would probably be slowly evolving and very dim.  

Here we have explored a very different scenario: mildly relativistic physical colllisions between two (initially) main sequence stars on circular EMRI orbits undergoing stable Roche-lobe overflow.  Focusing on the probably most common case of an inspiraling, low mass main sequence star interacting with an outspiraling brown dwarf, we have shown that at least a single collision between the stars is inevitable, when they occupy the same orbital phase at the line of nodes where their orbital planes cross (Fig.~\ref{fig:cartoon}).  

The initial collision is generally grazing.  Although only a tiny fraction of the stars' surfaces geometrically intersect, the enormous relative velocities ($\sim 0.1c$) cause massive shock heating of the stellar atmospheres.  When the mass of the SMBH $M_{\bullet} \lesssim 7 \times 10^{6}M_{\odot}$, this heating is insufficient to destroy the stars in a single encounter, and only a small fraction of the stellar mass is liberated, primarily from the more massive star.  This mass loss produces a gaseous accretion disk, which feeds the SMBH and causes a transient electromagnetic flare, potentially similar in appearance to observed candidate TDE flares.  
Mass loss from the collision also causes the orbital semi-major axis of the outer star to expand, separating it from the orbit of the inner star and delaying the next interaction for at least a decade and perhaps many millenia; the stars must wait for gravitational wave inspiral to realign their orbits before they can once again collide.  The net result of the ensusing string of grazing collisions is a ``death by a thousand cuts,'' producing a series of quasi-periodic accretion-powered flares, over a total duration of thousands of years or longer.  Conversely, if $M_\bullet \gtrsim 7 \times 10^6 M_\odot$, the first collision is likely powerful enough to completely destroy one or both stars; the ensuing flare will be more analogous to a classical TDE.

Although our estimates for the amount of shock heating and the resulting mass lost in grazing stellar collisions are crude, we expect that the qualitative features of the evolution described above should be qualitatively robust.  If the mass loss is smaller (larger) than we have assumed, this will simply increase (decrease) the total number of collisions before the stellar mass is eroded and decrease (increase) the interval between the collisions.  Future hydrodynamical simulations will better quantify the outcome of mildly relativistic stellar collisions and allow for a more accurate calibration of our model.  Future simulation work is also needed to quantify the fraction of the angular momentum of collisionally liberated gas which is fed back into the stellar orbit, accounting for the possible role of Lense-Thirring precession both on the disk and stellar orbits. 

Because the lifetime of the mass transfer evolution is comparable to the expected interval between EMRIs, colliding EMRI chains should occur at a rate comparable to the circular EMRI inspiral rate of $\gtrsim 10^{-7}$ yr$^{-1}$ per galaxy \citep{AmaroSeoane+12}.  Although this is still 2-3 orders of magnitude smaller than the predicted or observed TDE rate, we nevertheless conclude that collisional EMRIs can still contribute an appreciable fraction of the observed TDE rate, serving as ``TDE imposters."  This is because a given collisional interaction may produce a number $N_{\rm c} \sim 1-10^{4}$ of gas production events (eq.~\ref{eq:Ncoll2}) each of mass $\sim M_{\odot}/N_{\rm c} \sim 10^{-4}-1M_{\odot}$.  If accreted with high radiative efficiency, the luminosities produced after each collision could well explain those of many observed TDE flare candidates.  

Unlike in TDEs, the stellar debris from colliding EMRIs is tightly bound to the SMBH, allowing it to avoid the theoretically uncertain and perhaps lossy circularization process required to accrete highly eccentric TDE debris streams.  Our model provides a natural explanation for some flare light curves which appear to decay exponentially (e.g.~iPTF-16fnl;  \citealt{Blagorodnova+17}), or as power-laws shallower than $t^{-5/3}$,  depending on how efficiently the gas angular momentum liberated in the collision is fed back into the stellar orbit.  Viscous spreading of the gaseous disks produced by previous collisions in the chain can also provide a natural supply of radially extended, dense gas around the site of future collisions or TDEs, providing a possible medium for reprocessing the UV/X-ray accretion luminosity down to optical frequencies.

Future work is required to explore this new transient scenario in greater detail.  The rates of quasi-circular stellar EMRIs are quite uncertain, and hydrodynamic simulations are required to better understand mass loss in mildly relativistic stellar collisions.  They are also required to confirm whether Roche-overflowing stars can survive the substantial gas ablation expected during the many TDEs experienced between collisional interaction events ($\S\ref{sec:TDEstripping}$).  A population study of colliding EMRIs with a realistic distribution of initial stars and orbits would provide more accurate statistics on the expected range of outcomes.  Also deserving of future study is the role of stellar-mass black hole EMRIs, which should periodically pass through the main sequence EMRIs over the course of their mass-transfer evolution (Appendix \ref{sec:A}).

\acknowledgements

We thank Itai Linial, Cole Miller, and Re'em Sari for helpful comments on an early version of this manuscript.  We also thank Tiara Hung and Suvi Gezari for providing data on optical TDE light curves.  BDM gratefully acknowledges support from the National Science Foundation (AST-1410950, AST-1615084), NASA through the Astrophysics Theory Program (NNX16AB30G) and the Fermi Guest Investigator Program (NNX15AU77G, NNX16AR73G), the Research Corporation for Science Advancement Scialog Program (RCSA 23810), and the Alfred P.~Sloan Foundation.  Financial support was provided to NCS by NASA through Einstein Postdoctoral Fellowship Award Number PF5-160145.

\clearpage

\appendix

\section{Appendix A: Black Hole-Star Interactions}
\label{sec:A}

Due to mass segregation in galactic nuclei, EMRIs of stellar-mass black holes are probably more common than those of main sequence stars.  For this reason, black hole-star ``collisions'' could be more common than the star-star collisions focused on in this paper.  However, the effect of a BH piercing through a star which is parked in a mass-transferring orbit will likely be small.

First consider the energy imparted to the star by the tidal forces of the black hole.  The momentum and velocity change imparted as the black hole of mass $m_{\bullet}$ passes the star at a characteristic distance of $b \sim R_{\star}$ is estimated in the impulse approximation by the gravitational force times the fly-by time, $\Delta p = M_{\star} \Delta v \simeq (GM_{\star}m_{\bullet}/R_{\star}^{2})(R_{\star}/v_{\rm c})$.  The ratio of the deposited energy $\Delta E_{\rm t} \approx M_{\star}(\Delta v)^{2}/2$ to the gravitational binding energy of the star, $E_{\rm b} \approx GM_{\star}^{2}/R_{\star}$, is given by
\be
\frac{\Delta E_{\rm t}}{E_{\rm b}} \sim \left(\frac{m_{\bullet}}{M_{\star}}\right)^{2}\left(\frac{v_{\rm esc}}{v_{\rm c}}\right)^{2}, \label{eq:tidalImpulse}
\ee
where $v_{\rm esc} = (GM_{\star}/R_{\star})^{1/2} \approx 400$ km s$^{-1}$ is the escape speed from the star.  For typical parameters of $v_{\rm c} \sim 0.1-0.2 c$ (eq.~\ref{eq:vc}) and $m_{\bullet} \sim 10 M_{\star}$, this ratio is less than a percent.  Due to the higher mass of the black hole, it will migrate radially through the orbital radii occupied by the star star due via gravitational waves at a rate which is $\sim (m_{\bullet}/M_{\star})$ time faster than a star of mass $M_{\star}$, such that the number of close encounters $\sim N_{\rm GW}/N_{\rm coll}$ (eq.~\ref{eq:Nratio} for $M_{2} = m_{\bullet}$, $\chi = 1$) will typically be a modest $\sim 1-10$.  

The amount of stellar gas accreted by the black hole during the interaction will also be negligible.  While passing through the star, the black hole will accrete at a rate which is at most the Bondi-Hoyle rate $\dot{M}_{\rm B} = 4\pi G^{2}m_{\bullet}^{2}\rho_{\star}/v_{\rm c}^{3}$, where $\rho_{\star} \sim M_{\star}/(4\pi R_{\star}^{3}/3)$ is the typical stellar density encountered by the black hole.  The maximum mass accreted during a single passage of duration $t_{\rm c} \approx 2R_{\star}/v_{\rm c}$ is thus $M_{\rm acc} \approx \dot{M}_{\rm B}t_{\rm c}$, corresponding to a fraction of the star
\be
f_{\rm acc} = \frac{M_{\rm acc}}{M_{\star}} \approx 24\left(\frac{m_{\bullet}}{M_{\star}}\right)^{2}\left(\frac{v_{\rm esc}}{v_{\rm c}}\right)^{4}.
\ee
For typical parameters one finds $f_{\rm acc} \sim 10^{-5}-10^{-3}$, in which case even hundreds of collisions would not substantially erode the mass of the star.

Finally, we note that Eq. \ref{eq:tidalImpulse} predicts even weaker tidal perturbations in the canonical scenario of this paper, where the perturber is a low mass brown dwarf rather than a stellar mass BH.

%\section{Appendix}

\bibliographystyle{yahapj}
\bibliography{ms}
\end{document}